\documentclass[prx,reprint,amssymb,amsmath,longbibliography]{revtex4-1}
\usepackage[utf8x]{inputenc}
\usepackage{graphicx}
\usepackage{color}
\usepackage{colortbl}
\usepackage{grffile}
\usepackage[colorlinks,bookmarks=false,citecolor=blue,linkcolor=red,urlcolor=blue]{hyperref}
\usepackage{times}

\newcommand{\figref}[1]{FIG.~\ref{#1}}

\usepackage{multirow}
\usepackage{bm}

\newcommand{\eg}{e.\,g.}
\newcommand{\ie}{i.\,e.}
\newcommand{\wolog}{w.\,l.\,o.\,g.}
\newcommand{\dquote}[1]{``#1''}

\newcommand{\equref}[1]{Eq.~\eqref{#1}}
\newcommand{\refcite}[1]{Ref.~\cite{#1}}

\newcommand{\mathimg}[1]{
	\vcenter{\hbox{
		\includegraphics[scale=0.75]{#1}
	}}
}
\newcommand{\eqdefl}{:=}
\renewcommand{\vec}[1]{\bm{#1}}
\newcommand{\mat}[1]{\bm{#1}}

\usepackage{lipsum}

\definecolor{lightgray}{gray}{0.85}

\begin{document}
\title{Finite Correlation Length Scaling in Lorentz-Invariant Gapless iPEPS Wave Functions}

\author{Michael Rader}
\affiliation{Institut f\"ur Theoretische Physik, Universit\"at Innsbruck, A-6020 Innsbruck, Austria}
\author{Andreas M. L\"auchli}
\affiliation{Institut f\"ur Theoretische Physik, Universit\"at Innsbruck, A-6020 Innsbruck, Austria}
\begin{abstract}
It is an open question how well tensor network states in the form of an infinite projected entangled pair states (iPEPS) 
tensor network can approximate gapless quantum states of matter. Here we address this issue for two different physical scenarios: i) a 
conformally invariant  $(2+1)d$ quantum critical point in the incarnation of the transverse field Ising model on the
square lattice and ii) spontaneously broken continuous symmetries with gapless Goldstone modes exemplified by
the $S=1/2$ antiferromagnetic Heisenberg and XY models on the square lattice. We find that the energetically best wave functions display {\em finite} 
correlation lengths and we introduce a powerful finite correlation length scaling framework for the
analysis of such finite-$D$ iPEPS states. The framework is important i) to understand the mild limitations of the finite-$D$ 
iPEPS manifold in representing Lorentz-invariant, gapless many body quantum states and
ii) to put forward a practical scheme in which the finite correlation length $\xi(D)$ combined with field theory
inspired formulae can be used to extrapolate the data to infinite correlation length, i.e. to the
thermodynamic limit. The finite correlation length scaling framework opens the way for further exploration of quantum matter with
an (expected) Lorentz-invariant, massless low-energy description, with many applications ranging from condensed matter to high-energy physics.
\end{abstract}
\date{\today}
\maketitle
\tableofcontents

\section{Introduction}

The study of interacting quantum matter is of enormous current interest, with questions ranging from
quantum spin liquids, topological matter, correlated electrons in solids, ultracold atoms in 
optical lattices to strongly coupled quantum field theories. 

In this context numerical approaches play a very important role, with tensor networks being a central player. 
For problems in one spatial dimension
methods 
such as the density matrix 
renormalization group (DMRG)~\cite{White92,Schollwoeck05}, 
(infinite) matrix product states ((i)MPS)~\cite{Vidal2007,McCulloch2008,Schollwoeck11}, 
and the multiscale entanglement renormalization ansatz (MERA)~\cite{Vidal2008}
have proven to be very powerful, both for gapped states and for quantum critical states with a low-energy conformal field theory description~\cite{Tagliacozzo2008,Pollmann2009,Stojevic2015,Pfeifer2009}.

In two spatial dimensions infinite projected entangled pair states (iPEPS)~\cite{verstraete2004renormalization} have 
become a competitive numerical approach with successful applications to many problems in the field of quantum magnetism
and for strongly correlated fermions~\cite{Corboz2011,Corboz2012,Poilblanc2012,Corboz2014,Picot2016,Zheng2017}. Furthermore theoretical work has established how different forms of topological order can
be represented and understood in iPEPS wave functions with finite bond dimension~\cite{Schuch2012,Schuch2013}.

It is however an open question how well tensor network states in the form of an iPEPS 
tensor network can approximate gapless quantum states of matter. Here we address this issue for two different physical scenarios: i) a 
conformally invariant  $(2+1)d$ quantum critical point in the incarnation of the transverse field Ising model on the
square lattice and ii) spontaneously broken continuous symmetries with gapless Goldstone modes exemplified by
the $S=1/2$ antiferromagnetic Heisenberg and XY models on the square lattice. We find that the energetically best wave functions display {\em finite} 
correlation lengths and we introduce a powerful finite correlation length scaling framework for the
analysis of such finite-$D$ iPEPS states.

The outline of this paper is as follows: We start by providing a short introduction to the iPEPS framework and the energy optimization strategies
 in Sec.~\ref{sec:iPEPS}. In Sec.~\ref{sec:TFI} we study the critical properties of the $(2+1)d$ transverse field Ising model as an example of
a quantum critical point in the $(2+1)d$ Ising universality class. In Sec.~\ref{sec:symmbreaking} we present results for the $S=1/2$ antiferromagnetic
Heisenberg model and the $S=1/2$ XY model, as examples for continuous symmetry breaking. In Sec.~\ref{sec:discussion} we provide an extensive 
discussion and interpretation of the results obtained and we conclude in Sec.~\ref{sec:conclusions}.

\section{infinite (system) PEPS}
\label{sec:iPEPS}
Considering a two-dimensional quantum many-body system consisting of $p$-level particles,
placed on an infinite square lattice, one can make an ansatz for a wave function describing the system,
\begin{align}
	| \psi \rangle
	= \sum\limits_{\sigma_1, \sigma_2, \ldots}
	\mathimg{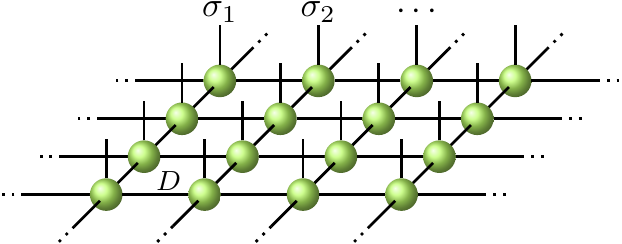}
	| \sigma_1, \sigma_2, \ldots \rangle
	\text{,}
	\label{eq:ipepsdef}
\end{align}
which is commonly known as {\em infinite system projected entangled pair state} (iPEPS) \cite{verstraete2004renormalization}.

The graph in \equref{eq:ipepsdef} is called a tensor network diagram, where nodes (edges) represent tensors (their corresponding indices). Edges connecting two tensors indicate summation indices and are of dimension $D$, which we call {\em bond dimension} of the iPEPS. The open indices $\sigma_1$, $\sigma_2$, \ldots are of dimension $p$.
In this work we only consider iPEPS with a single-site unit cell, \ie{} iPEPS where the same tensor is used for each site, but all the techniques described here can be generalized to arbitrary unit cells.

iPEPS are a straight-forward generalization of infinite system matrix product states (iMPS) for two spatial dimensions and obey an area law for the entanglement entropy as well \cite{verstraete2004renormalization,eisert2010}.
In contrast to iMPS, iPEPS can be constructed to have infinite correlation lengths, already for the simplest non-trivial case, $D=2$~\cite{Verstraete2006}.

\subsection{Contraction}
The key challenge in all iPEPS algorithms is the so-called contraction of the state.
For instance, to evaluate single-site observables, one needs the single-site density matrix,
\begin{align}
	\rho_1 =
	\mathimg{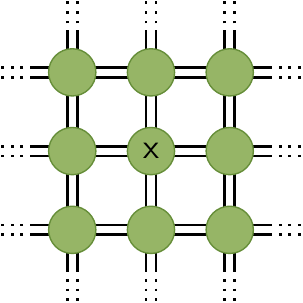}
	\text{,}
\end{align}
which consists of an infinite sum of double-layer tensors,
\begin{align}
	\mathimg{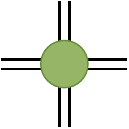}
	\eqdefl
	\mathimg{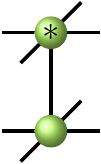}
	\quad
	\text{and}
	\quad
	\mathimg{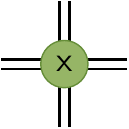}
	\eqdefl
	\mathimg{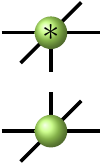}
	\text{.}
\end{align}
The idea of a contraction is to find an approximation with controllable error for this infinitely large tensor network.
There are several ways to go: Finding an approximate environment in the form of a boundary matrix product state (bMPS) \cite{Jordan2008} or a corner transfer matrix (CTM) environment \cite{Orus2009} or by directly applying renormalization group (RG) schemes such as tensor RG (TRG) \cite{levin2007trg}, tensor-entanglement RG (TERG) \cite{gu2008terg}, second RG (SRG) \cite{xie2009srg}, higher-order TRG (HOTRG) \cite{xie2012hotrg} or tensor network renormalization (TNR) \cite{evenbly2015tnr}.

Formally, a bMPS is nothing but an iMPS, with bond dimension $\chi$, that is an approximation for the dominant eigenvector of a transfer matrix of double layer iPEPS tensors,
\begin{align}
	\mathimg{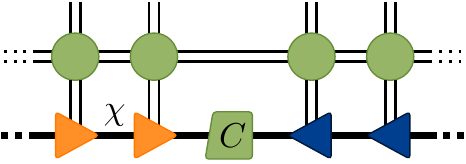}
	\approx
	\mathimg{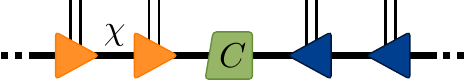}
	\text{,}
\end{align}
with corresponding eigenvalue $1$.
Note that the quality of the approximation can be improved by increasing $\chi$.
From the singular values, $\zeta_j$, of the matrix $C$ (Rényi) entropies,
\begin{align}
	S_\mathrm{bMPS}^{(\alpha)} = \frac{1}{1-\alpha} \log \left( \sum\limits_{j} \zeta_j^\alpha \right)
	\text{,}
\end{align}
can be computed.
In addition to the bMPS tensors one has to determine a horizontal dominant eigenvector,
\begin{align}
	\mathimg{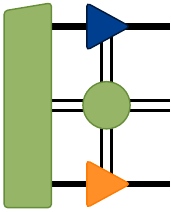}
	=
	\mathimg{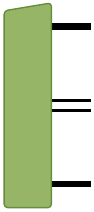}
	\text{,}
\end{align}
with corresponding eigenvalue $1$
to be able to write the single-site density matrix as
\begin{align}
	\rho_1 \approx
	\mathimg{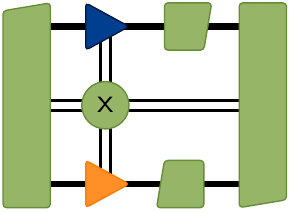}
	\text{.}
\end{align}
The state-of-the-art method to find such a bMPS is described in \refcite{fishman2017faster} and has also been used in this work.

Another way to contract a state, is to find corner transfer matrices (CTMs),
\begin{align}
	\mathimg{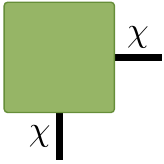}
	\approx
	\mathimg{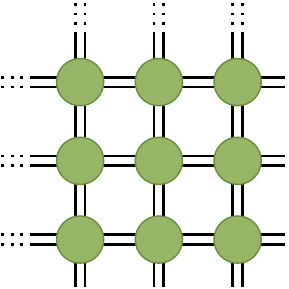}
	\text{.}
\end{align}
and half-line transfer tensors (HLTTs),
\begin{align}
	\mathimg{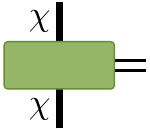}
	\approx
	\mathimg{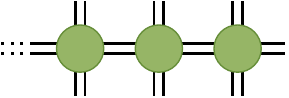}
	\text{,}
\end{align}
where again the contraction dimension $\chi$ is used to control the error of the approximation.
With the CTMs and HLTTs, the single-site density matrix can be written as
\begin{align}
	\rho_1
	\approx
	\mathimg{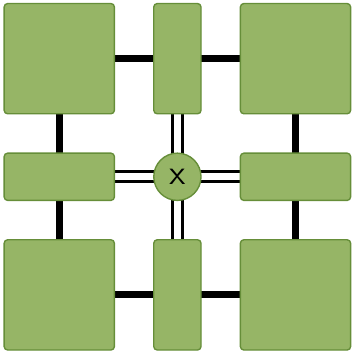}
	\text{.}
\end{align}
A powerful numerical tool to find CTMs and HLTTs is the so-called CTM renormalization group (CTMRG) \cite{Orus2009}. 
The specific CTMRG procedure introduced in \refcite{corboz2014competing} has been used in this work, as it is a particularly stable variant of CTMRG.

For this work both bMPS and CTMRG contractions have been implemented and we observe that for equal $\chi$ both methods give almost identical results.

\subsection{Energy Optimization}
As iPEPS are especially well-suited to describe ground states, energy optimization algorithms for iPEPS are of particular interest.
For almost a decade there was only a single strategy for this task: imaginary time evolution~\cite{Jordan2008} in various variants.
Recently Refs.~\cite{Corboz2016b} and~\cite{Vanderstraeten2016} introduced new direct variational approaches, which achieved lower energies than imaginary time evolutions, even in the limit of vanishing Trotter step size. Both variational methods rely on interaction contractions, similar to the norm contractions described in the previous subsection, but including all interaction terms of the Hamiltonian.

The first method \cite{Corboz2016b} makes use of CTMRG interaction contractions to formulate a generalized eigenvalue problem (GEVP) for a given iPEPS, where the eigenvector corresponding to the lowest eigenvalue is used to propose the next iPEPS tensor in the minimization run.

The second method  \cite{Vanderstraeten2016} computes the energy gradient of an iPEPS from a bMPS interaction contraction, such that any gradient minimization technique can be used, \eg{} conjugate gradient (CG) methods or quasi-newton methods such as the Broyden--Fletcher--Goldfarb--Shanno (BFGS) algorithm \cite{broyden1970,fletcher1970,goldfarb1970,shanno1970}.

However, it should be noted that also CTMRG interaction contractions can be used to obtain energy gradients and vice versa bMPS interaction contractions to obtain the optimization GEVP~\cite{Rader_Unpublished}.

For iPEPS energy optimizations in this work bMPS interaction contractions in combination with the BFGS algorithm were primarily used, as this method turned out to be the most stable one.
Also, the BFGS algorithm seems to be more stable close to the minimum compared to CG methods.
Some states also have been optimized using a brute force minimization method, \ie{} by computing finite difference energy gradients.
All iPEPS optimized in this work have a single-site unit cell with a complex tensor which was forced to be invariant under spatial rotations. No symmetries
at the virtual level were imposed.
It turned out that starting with several random states and small contraction dimensions ($16 \lesssim \chi \lesssim 32$) is the most economic way to bootstrap energy minimizations. The contraction dimension is then successively increased -- in our optimizations up to values of $\chi = 512$.
For the transverse field Ising model we observed that it can also be beneficial to use intermediate minimization results to initialize minimizations for nearby values of the transverse field.

\subsection{Correlation Length}
\label{sec:ipepscorrelationlength}
A correlation function, $c(r)$, of two arbitrary observables can be written as
\begin{align}
	c(r) = 
	\mathimg{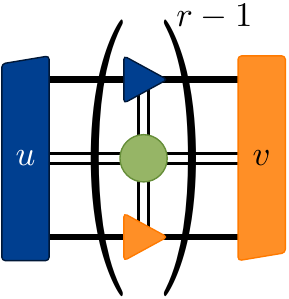}
	= \vec{u}^T \mat{A}^{r-1} \vec{v}
	\text{,}
\end{align}
using a bMPS contraction, where the tensors $\vec{u}$ and $\vec{v}$ contain the specific observables.
It should be noted that the same correlation function can be expressed by replacing the bMPS tensors with HLTTs from a CTMRG contraction.

We write the eigendecomposition of the transfer matrix as
\begin{align}
	\mathimg{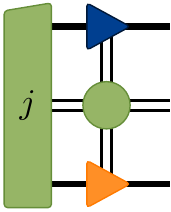}
	=
	\lambda_j
	\mathimg{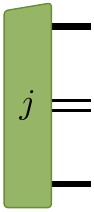}
	\text{,}
\end{align}
with $|\lambda_0| \geq |\lambda_1| \geq \ldots$ and \wolog{} we assume $\lambda_0 = 1$.
Then the dominant correlation length of the system, $\xi$, can be extracted as
\begin{align}
	\xi = -\frac{1}{\log | \lambda_1 | }
	\text{.}
\end{align}
With
this method the dominant correlation length can be extracted without even knowing which observables lead to the corresponding correlation function.

It should be noted that in contrast to local observables, the correlation length requires huge contraction dimensions $\chi$ to converge.
Incorporating the ideas of \refcite{rams2018precise} we observe the functional behavior
\begin{align}
	\frac{1}{\xi(\chi)}
	= \frac{1}{\xi(\infty)} + k \log \left| \frac{\lambda_1(\chi)}{\lambda_{2}(\chi)} \right|
	\text{,}
	\label{eq:xiscaling}
\end{align}
which enables to extract a converged value for the correlation length, $\xi(\infty)$, precisely already for moderate values of $\chi$. In cases where $\lambda_1 \simeq \lambda_2$ (due to degeneracy, as observed e.g.~in the $S=1/2$ Heisenberg model) one should use the largest eigenvalue different from  $\lambda_1$ instead of $\lambda_2$ for the scaling in \equref{eq:xiscaling}.

\section{Quantum Critical Behavior in a $(2+1)d$ Conformal Field Theory}
\label{sec:TFI}

\begin{figure}
 \includegraphics[width=0.7\linewidth]{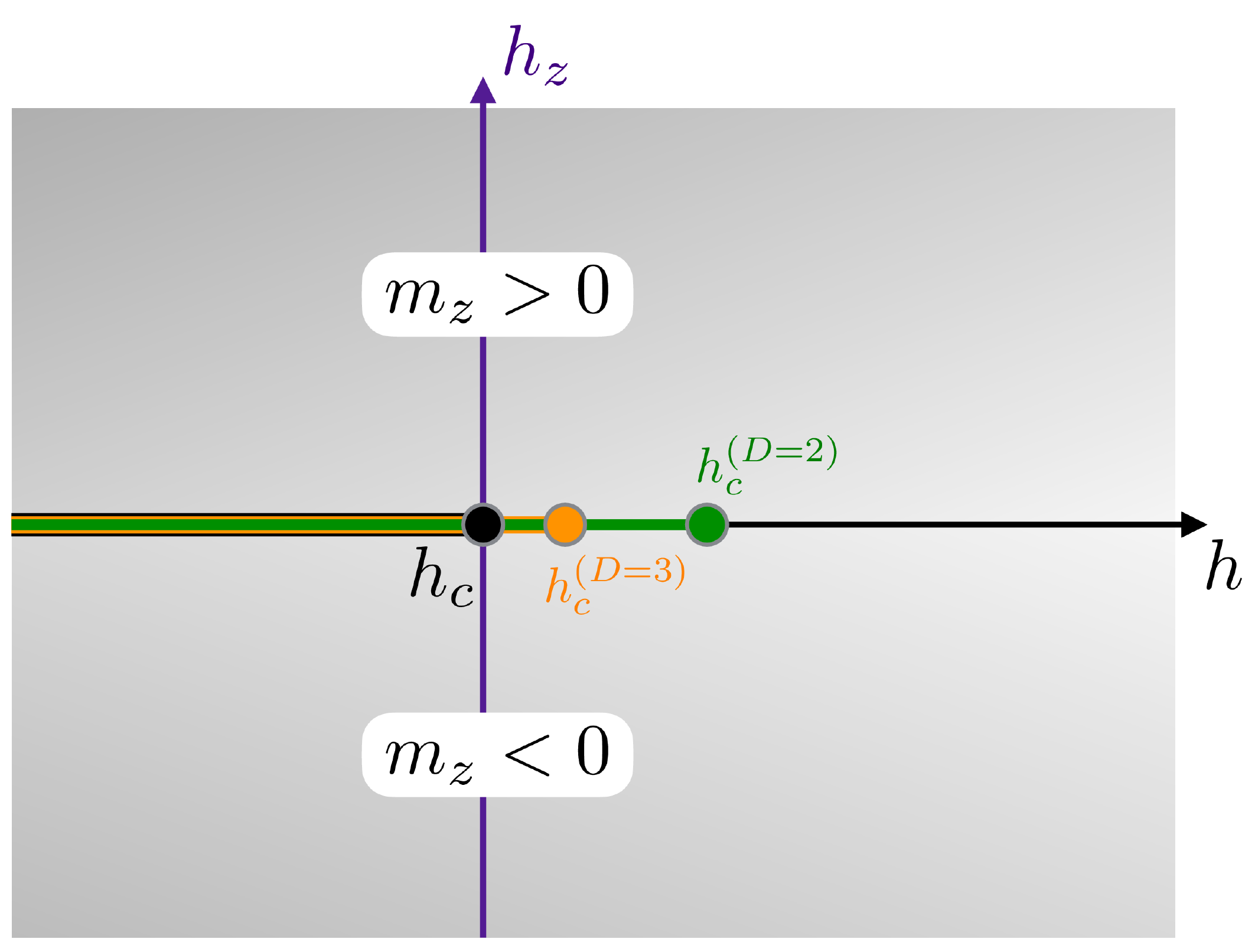}
 \caption{\label{fig:Ising_PhaseDiagram} Phase diagram of the ferromagnetic ($J=1$) transverse field Ising model 
 with an additional longitudinal field $h_z$. The thick, colored lines for $h<h_c$ or $h<h_c^{(D)}$ highlight the horizontal extent of the 
 spontaneous symmetry breaking line at $h_z=0$.}
\end{figure}

In a first application we study the critical behavior of our variationally optimized iPEPS tensors
for a quantum critical point in the $(2+1)d$ Ising universality class. In the following we call this universality class described
by a $(2+1)d$ dimensional, Lorentz-invariant conformal field theory (CFT) the {\em 3d Ising CFT}. 
Note that this critical behavior is distinct from the one observed in the so called {\em Ising PEPS}~\cite{Verstraete2006}, which amounts to
promoting the thermal partition function of the 2d classical Ising model into a two-dimensional quantum many body wave function in PEPS form
with bond dimension $D=2$.
This wave function can be parametrized by the temperature $T$ entering the partition function, and at $T=T_c$ describes a
critical wave function with algebraically decaying correlation functions. However the critical properties of this wave function are 
described by the {\em 2d Ising CFT}. 

\subsection{Overview}

\begin{table}[b]
\begin{tabular}{|c|c|c|c|}
\hline
exponent 			    & $d=2$ 	& $d=3$			& $d=4$\\
\hline
\hline
$\Delta_\sigma$   & $1/8$		&  $0.518\ 148\ 9(10)$  	& $1$\\
$\Delta_\epsilon$ & $1$		&  $1.412\ 625(10)$		& $2$\\
\hline
\hline

$\nu_\sigma=1/(d-\Delta_\sigma)$    					&  $8/15$	& $0.402\ 925\ 0(2)$		& $1/3$	\\
$\nu_\epsilon=1/(d-\Delta_\epsilon)$    				& $1$	& $0.629\ 970(4)$		&$1/2$\\
$\alpha_\sigma=\Delta_\epsilon \times \nu_\sigma $   	& $8/15$	& $0.569\ 182(4)$		& $2/3$\\
$\alpha_\epsilon=\Delta_\epsilon \times \nu_\epsilon$ 	& $1$	& $0.889\ 91(1)$		& $1$\\
$\beta_\sigma = \Delta_\sigma \times \nu_\sigma$    		& $1/15$	& $0.208\ 775\ 1(5)$		& $1/3$\\
$\beta_\epsilon = \Delta_\sigma \times \nu_\epsilon$   	& $1/8$	& $0.326\ 418(2)$		& $1/2$\\
\hline
\end{tabular}
\caption{Upper two rows: relevant scaling dimensions of the Ising CFT in $d=2,3,4$  space-time dimensions, results 
for $d=3$ are from the most recent conformal bootstrap study~\cite{Kos2016}. The lower six
rows denote the additional critical exponents probed in this work, which are derived from the scaling dimensions above.}
\label{tab:IsingCFT}
\end{table}

\begin{figure*}
 \includegraphics[width=0.8\linewidth]{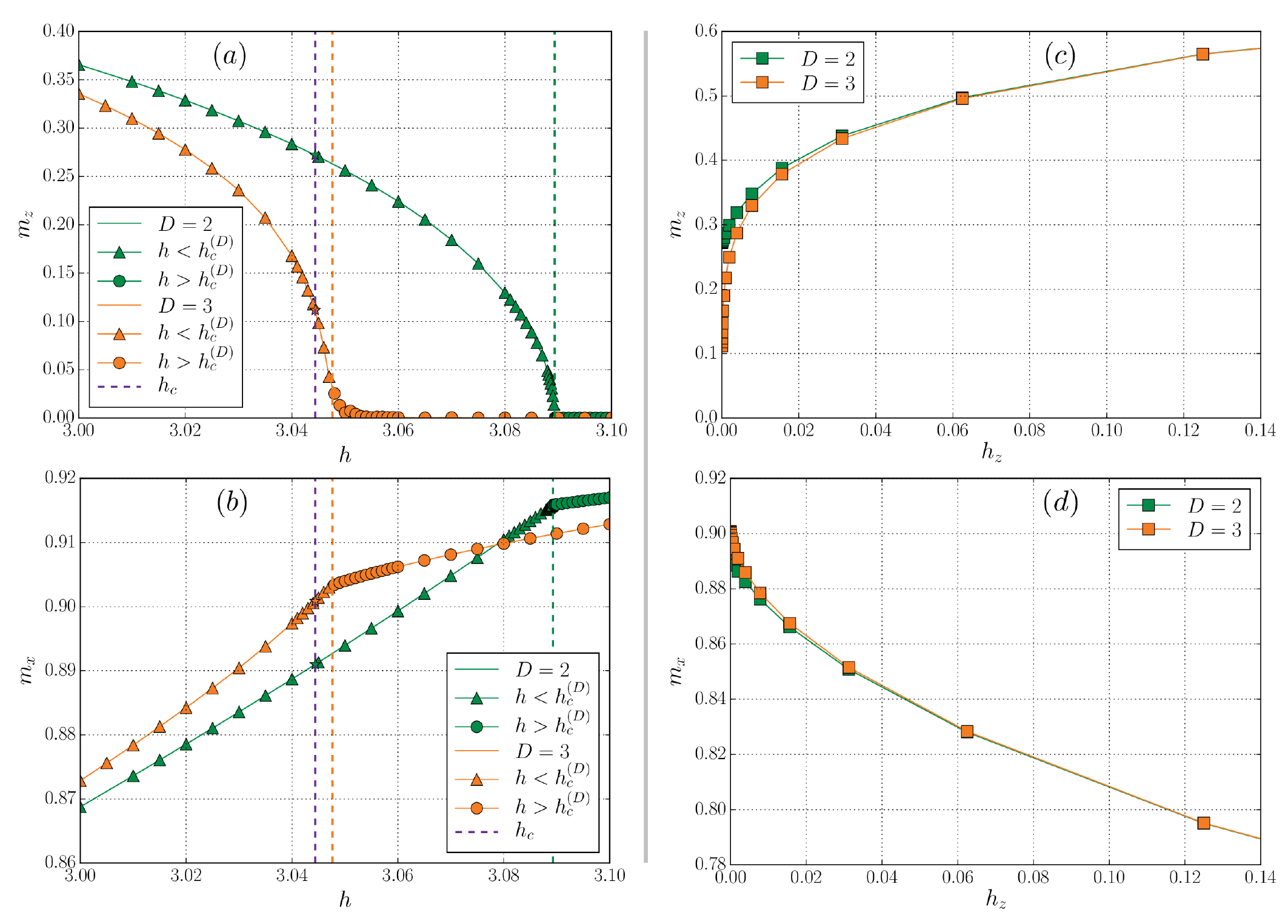}
 \caption{\label{fig:iPEPS_TFI_hx_hz} iPEPS results for the magnetizations per site $m_z=|\langle\sigma_i^z\rangle|$ and $m_x=\langle\sigma_i^x\rangle$
 as a function of $h$ ($h_z=0$) in panels (a) and (b) and as a function of $h_z$ ($h=h_c$) in panels (c) and (d). Result for iPEPS bond dimension $D=2,\ 3$
 are shown. The vertical dashed lines indicate the critical values of $h_c^{(D)}$ for both $D$ values, as well as the literature reference for $h_c$~\cite{Bloete2002}.}
\end{figure*}

The Hamiltonian studied in this section is the transverse field Ising model on an infinite square lattice 
with an additional longitudinal magnetic field,
\begin{equation}
H_\mathrm{TFI}=-J \sum_{\langle i,j\rangle} \sigma_i^z\sigma_j^z -h\sum_i \sigma_i^x -h_z\sum_i \sigma_i^z\ ,
\label{eq:Ham_TFI}
\end{equation}
where $J=1$ denotes the strength of the ferromagnetic Ising interactions and sets the energy scale, $h$ denotes the transverse field 
while $h_z$ parametrizes the longitudinal magnetic field. The phase diagram of this model is sketched in Fig.~\ref{fig:Ising_PhaseDiagram}. For $h_z=0$ the
model has a $\mathbb{Z}_2$ spin inversion symmetry and features a quantum phase transition at $h/J=h_c=3.04438(2)$~\cite{Bloete2002}, which separates a paramagnetic
phase with $m_z\equiv \langle \sigma^z_i \rangle =0$ for $h>h_c$ from a symmetry broken phase with $m_z\neq 0$ for $h<h_c$. The quantum critical point
at $h=h_c, h_z=0$ is 
described
by the 3d Ising CFT. For all finite $h_z\neq0$ there is {\em no} critical behavior and the $z$-magnetization
$m_z$ is finite as a response to the finite $h_z$. In the entire phase diagram with $h\neq0$ the transverse $x$-magnetization, $m_x\equiv \langle \sigma_i^x\rangle$,
is finite.

In the following we will explore the physics in the vicinity of the critical point $h=h_c, h_z=0$
using iPEPS simulations. In order to provide field theoretical predictions to compare with we 
briefly review the properties of the 3d Ising CFT for our purposes. The 3d Ising CFT has two 
relevant perturbations, the $O_\sigma$ and the $O_\epsilon$ field. Their scaling dimensions $\Delta_\sigma$
and $\Delta_\epsilon$ are given in Tab.~\ref{tab:IsingCFT}.
They clearly differ in $2d$ and $3d$, which leads to distinct critical behavior.
In the transverse field Ising model it is expected that
in the continuum limit we can identify $\sigma^z_i \sim O_\sigma(\bm{r}_i)$, while $\sigma^x_i \sim O_\epsilon(\bm{r}_i) + \mathrm{const}$.

We perturb a general CFT in $d$ space-time dimensions with a relevant perturbation $\phi$ with scaling dimension $\Delta_\phi$:
$$ \mathcal{H}(\lambda)= \mathcal{H}_\mathrm{CFT} + \lambda \int \mathrm{d}x^{d-1}\ \hat{\phi}.$$
Since the perturbation is relevant, i.e.~$\Delta_\phi <d$, it opens a mass gap proportional to $\lambda^\frac{1}{d-\Delta_\phi}$, 
respectively it leads to a {\em finite} correlation length $\xi \sim \lambda^{-1/(d-\Delta_\phi)}\equiv \lambda^{-\nu_\phi}$. In the transverse
field Ising model it is understood that the microscopic coupling $(h-h_c)$ couples to the field $O_\epsilon$, while $h_z$ couples to the field
$O_\sigma$.

The 3d Ising CFT has two relevant perturbations, which
translate into the two correlation length exponents,
$\nu_\epsilon=1/(d-\Delta_\epsilon)$
and
$\nu_\sigma=1/(d-\Delta_\sigma)$,
which appear in the scaling relations
$\xi \sim |h-h_c|^{-\nu_\epsilon}$
and
$\xi \sim |h_z|^{-\nu_\sigma}$.

We are in a situation where the perturbed theory can display magnetizations,
and we thus define critical exponents $\alpha_\phi,\beta_\phi$ which describe how 
the magnetizations grow as a function of the perturbing coupling for a perturbation $\phi$.
$\alpha_\phi$ is the exponent we use when measuring $ m_{x,c} - \langle \sigma^x_i\rangle\sim \langle O_\epsilon\rangle$, while we use $\beta_\phi$ 
for the $\langle \sigma^z_i \rangle\sim \langle O_\sigma\rangle$ observable. The subscript $\sigma,\epsilon$ denotes the perturbing field
as for the correlation length exponents before. For the transverse magnetization we expect $|m_{x,c}-m_x| \sim |\lambda|^{\alpha_\phi}$,
while for the longitudinal magnetization we expect $m_z\sim |\lambda|^{\beta_\phi}$. The definitions and values for these exponents can be found in
Tab.~\ref{tab:IsingCFT}. Some of the exponents we have introduced here for clarity reasons are also more commonly
known as $\beta\equiv\beta_\epsilon$, $\delta\equiv 1/\beta_\sigma$ and $\nu\equiv\nu_\epsilon$ in the statistical mechanics
literature. Note that the $d=4$ critical exponents are equivalent to the mean field exponents. This is because $d=4$ is the upper 
critical dimension for the Ising criticality, i.e.~the dimension where mean field theory becomes exact.

\begin{figure*}
 \includegraphics[width=0.8\linewidth]{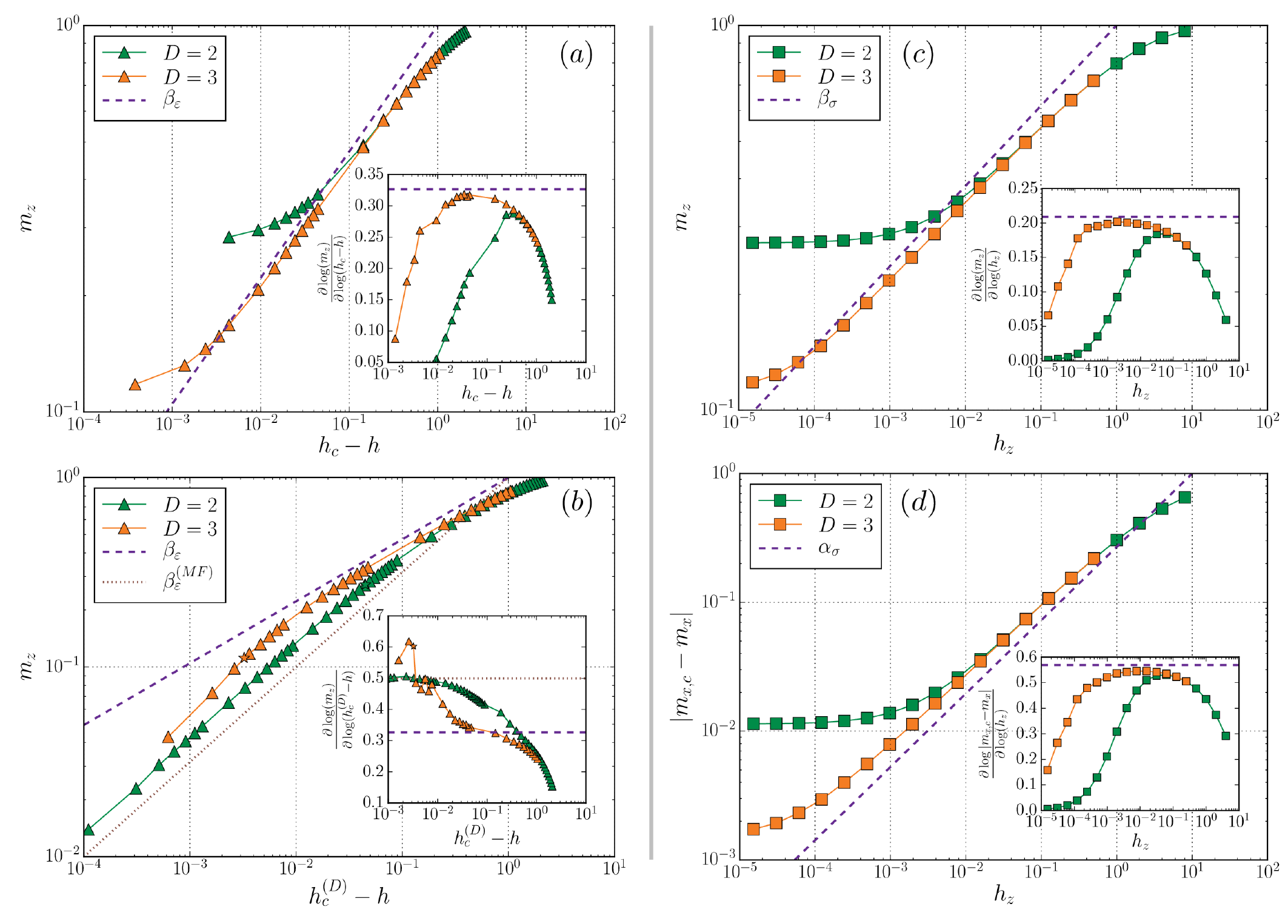}
 \caption{\label{fig:iPEPS_TFI_criticality} Analysis of the critical behavior of the $m_z$ and $m_x$ magnetizations in the vicinity of the critical point
 at $h=h_c,h_z=0$. (a) $m_z$ as a function of $h_c-h$, inset shows good convergence to the expected $\beta_\epsilon$ exponent. (b) $m_z$ as a function of $h_c^{(D)}-h$,
 inset shows crossover to mean field behavior. (c) $m_z$ as a function of $h_z$, inset shows good convergence to the expected $\beta_\sigma$ exponent.
 (d) $|m_x-m_{x,c}|$ as a function of $h_z$, inset shows good convergence to the expected $\alpha_\sigma$ exponent. The expected exponents are tabulated in Tab.~\ref{tab:IsingCFT}.
}
\end{figure*}

\subsection{iPEPS Results for the Transverse Field Ising Model}

The transverse field Ising model (with $h_z=0$) has been studied extensively with both finite-size PEPS and iPEPS approaches
in the past~\cite{Jordan2008,Gu2008,Orus2009,Orus2012,Liu2010,Lubasch2014,Phien2015a,Phien2015b,Vanderstraeten2016}.
A common feature of all the these simulations is that for $h$ below a bond dimension $D$ dependent value, $h_c^{(D)}$, the 
system shows a finite $z$-magnetization $m_z$. However in the past the values of $h_c^{(D)}$ and the functional behavior of $m_z$ in its
vicinity did depend significantly on the tensor optimization methodology. We believe the newest generation of optimization algorithms
put forward in Refs.~\cite{Corboz2016b,Vanderstraeten2016} do not suffer from these shortcomings anymore, so that a detailed analysis
of the {\em intrinsic} iPEPS  finite-$D$ behavior close to the 3d Ising CFT is finally possible.

\subsubsection{Magnetizations in the vicinity of the critical point}

We start by presenting in Fig.~\ref{fig:iPEPS_TFI_hx_hz} the behavior of the $z$- and the $x$-magnetization 
at $h_z=0$
by varying $h$ in panels (a) and (b) for bond dimensions $D=2,3$. As previously reported we also observe a $D$-dependent value $h_c^{(D)}$
where $m_z$ vanishes, while $m_x$ displays a kink. We find $h_c^{(D=2)}\approx 3.0893$, while $h_c^{(D=3)}\approx 3.0476$. 
Note that the $D=3$ result  differs only by about one part per thousand from the reference critical value $h_c=3.04438(2)$~\cite{Bloete2002}. In an earlier
study based on one-dimensional iMPS states for the $(1+1)d$ transverse field Ising model, a bond dimension $D>10$ was required to reach a comparable accuracy~\cite{Liu2010}.
We have also tried to optimize $D=4$ tensors, but albeit technically possible, it turns out to be extremely difficult to obtain energies which are lower than our best
$D=3$ results so far. We will come back to this observation later in this section. Finally in panels (c) and (d) we display $m_z$ and $m_x$ 
along an orthogonal cut at fixed $h=h_c$ with varying $h_z>0$, i.e.~along the violet axis in Fig.~\ref{fig:Ising_PhaseDiagram}. 

While the plots in Fig.~\ref{fig:iPEPS_TFI_hx_hz} seem to suggest large differences between $D=2$ and $D=3$ it should be noted that the $h$ and $h_z$ ranges
shown are already quite small. Shown on a scale $h\in[0,4]$ it would be difficult to visually spot the differences between the two $D$ values.

\subsubsection{Critical exponents}

In a next step we explore the critical behavior contained in the presented data. In Fig.~\ref{fig:iPEPS_TFI_criticality}(a) we plot $m_z$ as a function of $h_c-h$
on a $\log$-$\log$ scale. For comparison we plot a straight line with the expected slope $\beta_\epsilon$ as a guide to the eye. In the corresponding inset we numerically
calculate the derivative and find a collapse between the $D=2$ and $D=3$ data at larger distances from the critical point. The $D=2$ running estimate for the critical exponent reaches a maximum of $\sim 0.29$ and then drops to zero as $h_c-h$ goes to zero.  The $D=3$ running estimate rises to $\sim 0.32$ before it also drops to zero 
as $h_c-h$ goes to zero. We expect that $D>3$ would get even closer to the expected 
$\beta_\epsilon\approx0.326\ 418(2)$, before dropping to zero as $h_c-h$ goes to zero. The drop to zero is clearly due to the finite-$D$ remnant $m_z$ at the thermodynamic value
$h_c$. Let us therefore investigate what happens when we analyze the data as a function of $h_c^{(D)}-h$ instead. Panel (b) of Fig.~\ref{fig:iPEPS_TFI_criticality} displays the corresponding
data. The $D=2$ data clearly shows a limiting mean-field behavior $\beta_\epsilon^{(MF)}=1/2$ at small $h_c^{(D)}-h$ (in the inset), as observed previously in Ref.~\cite{Liu2010}.
The $D=3$ data shows some hint of an intermediate plateau around the true 3d Ising CFT value for $\beta_\epsilon$ before also crossing over to the mean-field value $\beta_\epsilon^{(MF)}$ at small $h_c^{(D)}-h$. The analysis of the $m_x$ magnetization is less clean and shown in Fig.~\ref{fig:TFI_mx_h} in App.~\ref{app:mx_h}. In Panels (c) and
(d) of Fig.~\ref{fig:iPEPS_TFI_criticality} we present the analogous analysis for both $m_z$ and $m_x$ when staying at $h=h_c$ while tuning $h_z>0$. The numerical derivatives 
provide running averages for $\beta_\sigma$ and $\alpha_\sigma$ which converge nicely towards to the expected values as we increase $D$ from $2$ to $3$. The maximal values for $D=3$
are only a few percent below the 3d Ising CFT results.

So we learn that when the location of the critical point is known beforehand, the critical behavior can be determined quite accurately already with a surprisingly small
bond dimension of $D=3$. In the vicinity
of the finite $D$ critical points we however observe mean field behavior and due to the crossover it is more difficult to extract the genuine critical behavior.

\begin{figure*}
  \includegraphics[width=0.95\linewidth]{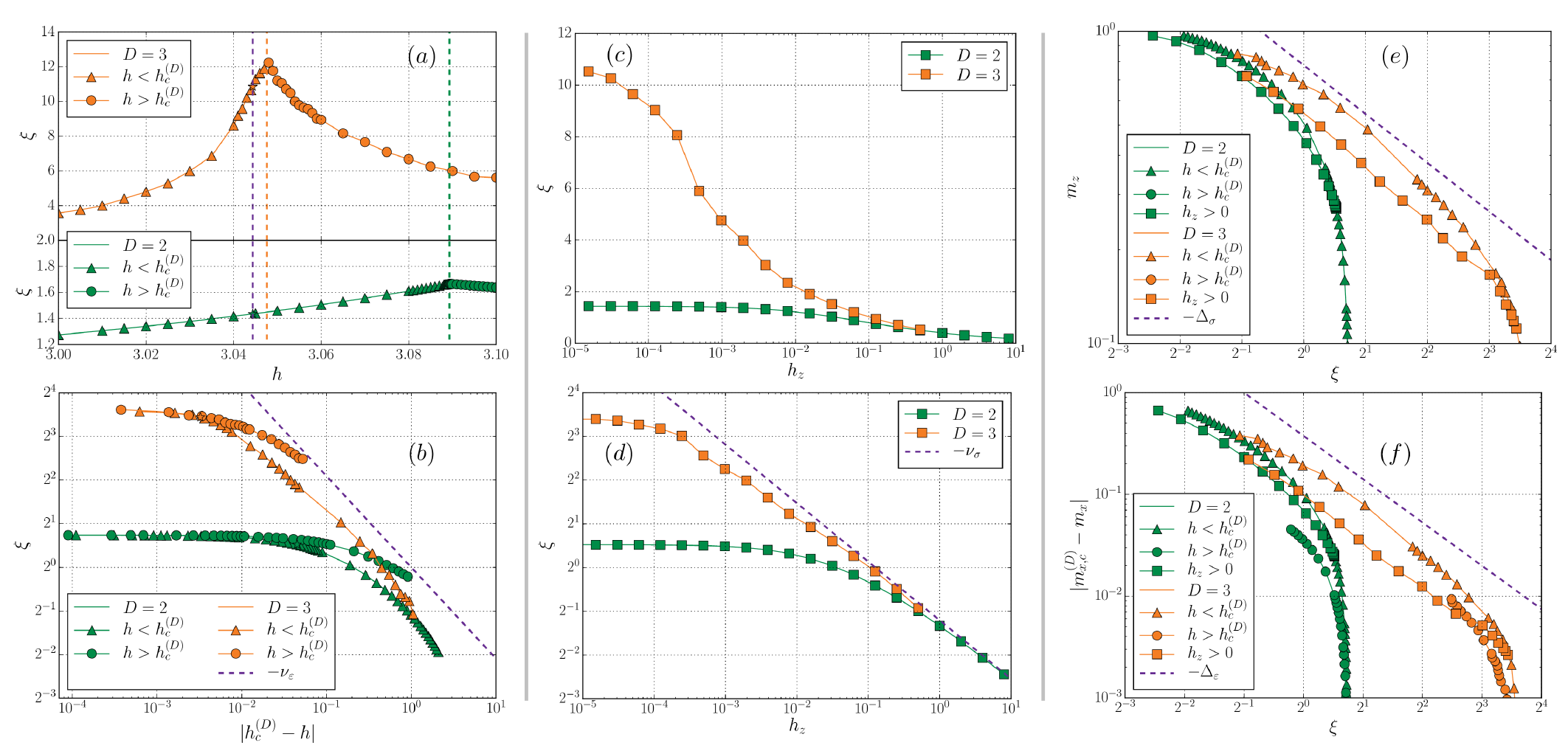}
 \caption{\label{fig:iPEPS_TFI_corrlength}  Correlation lengths of the variationally optimized iPEPS states for $D=2$ and $D=3$. 
 (a) correlation lengths as a function of the transverse field $h$. (b) correlation lengths as function of $|h_c^{(D)}-h|$ on a $\log$-$\log$ scale,
 including the theoretically expected slope $-\nu_\epsilon$.
 (c) correlation lengths as a function of the longitudinal field $|h_z|$. (d) correlation lengths as function of $|h_z|$ on a $\log-\log$ scale,
 including the theoretically expected slope $-\nu_\sigma$.
 (e) and (f) display the expectation value of $m_z$ and $|m_{x,c}^{(D)}-m_x|$ respectively as a function of the correlation length $\xi$, on a 
 logarithmic scale. The expected slopes in the two cases are directly the (negated) scaling dimensions $-\Delta_\sigma$ and $-\Delta_\epsilon$. 
 The values of the exponents and scaling dimensions can be found in Tab.~\ref{tab:IsingCFT}.
 }
\end{figure*}

\subsubsection{Correlation lengths}

After having analyzed the critical behavior of local observables as a function of perturbing couplings, we
now investigate the correlation lengths in our optimized iPEPS wave functions in the vicinity of the critical 
point. In the vicinity and at a quantum critical point in $(1+1)d$ represented with a finite bond dimension iMPS
we know that only a {\em finite} correlation length can appear. Since iPEPS is in principle able to represent
wave functions with algebraically decaying correlations, i.e.~states with infinite correlation lengths~\cite{Verstraete2006}, 
it is not a priori clear what to expect in our optimized iPEPS wave functions. Let us note first that the correlation 
lengths for $D=1$ (product states) vanish identically, even though the spontaneous magnetization shows 
critical mean-field behavior at $h_c^{(D=1)}=4$ (not shown). Based on the technology presented in subsection~\ref{sec:ipepscorrelationlength}
we have determined the largest correlation lengths for $D=2$ and $D=3$ iPEPS states along the previously investigated
cuts in the $(h,h_z)$ plane. The results are shown in panels (a) and (c) of Fig.~\ref{fig:iPEPS_TFI_corrlength}. The observed 
correlation lengths for $D=2$ do not grow beyond $\xi^{(D=2)}_\mathrm{max} \approx 1.67$ lattice spacings, and they reach their maximum at $h=h_c^{(D=2)}, h_z=0$.
We are quite confident that this short correlation length is not an artefact of incomplete optimization, but is a genuine feature of
optimized, translationally invariant, finite-$D$ iPEPS wave functions for Lorentz-invariant quantum critical systems with a 3d space-time description. 
For $D=3$ we also observe finite correlation lengths, but now the maximum is substantially larger: $\xi^{(D=3)}_\mathrm{max} \approx 12.2$ at $h=h_c^{(D=3)}, h_z=0$
~\footnote{Since the energy optimization of states with such a large correlation length is computationally quite demanding, there is a small uncertainty in the value of $\xi^{(D=3)}_\mathrm{max}$.}. So both $D$ values seem to indicate that our variational optima feature a {\em finite} correlation length. This is one of the key results of this paper, 
whose possible origin we are going to discuss later. We will however show in the following, that the finite correlation length is also a blessing, as it helps us to understand and organize the finite $D$ effects in field theoretically motivated formulas based on $\xi(D)$. 

Before doing this, let us investigate the functional behavior of the correlation length in the
vicinity of the critical point. Depending on the cut in parameter space we expect $\xi \sim |h_c-h|^{-\nu_\epsilon}$ or $\xi \sim |h_z|^{-\nu_\sigma}$,
with the values of $\nu_\epsilon$ and $\nu_\sigma$ given in Tab.~\ref{tab:IsingCFT}. Indeed the data in panels (b) and (d) of Fig.~\ref{fig:iPEPS_TFI_corrlength} shown on
a $\log$-$\log$ scale are (roughly) compatible with the expected correlation length exponents in some intermediate window of the couplings. This is expected since far away from
the critical point we are outside the quantum critical regime, while very close to the critical point $\xi$ saturates. Still the agreement for the $h_z$-detuning is much better than 
the $|h_c-h|$-detuning.

Finally we plot the expectation values of the two magnetizations $m_z$ and $|m_{x,c}-m_x|$ as a function of the measured correlation length $\xi$ (for both parameter cuts) 
in panels (e) and (f) of Fig.~\ref{fig:iPEPS_TFI_corrlength}. As discussed earlier we expect this relation to be governed by the scaling dimensions $\Delta_\sigma$
and $\Delta_\epsilon$ respectively. While the $D=2$ results do not match too well, the $D=3$ results for both parameter cuts and both observables are in good agreement with the expected
scaling dimensions.

Even though it seems that the critical exponents and scaling dimensions can be obtained more precisely based on the observables as a function of the couplings than of
the correlation lengths, it is nevertheless
rewarding to observe that the correlation lengths are also following the expected quantum critical behavior with increasing $D$, within the stated limitations.

\begin{figure}
 \includegraphics[width=0.95\linewidth]{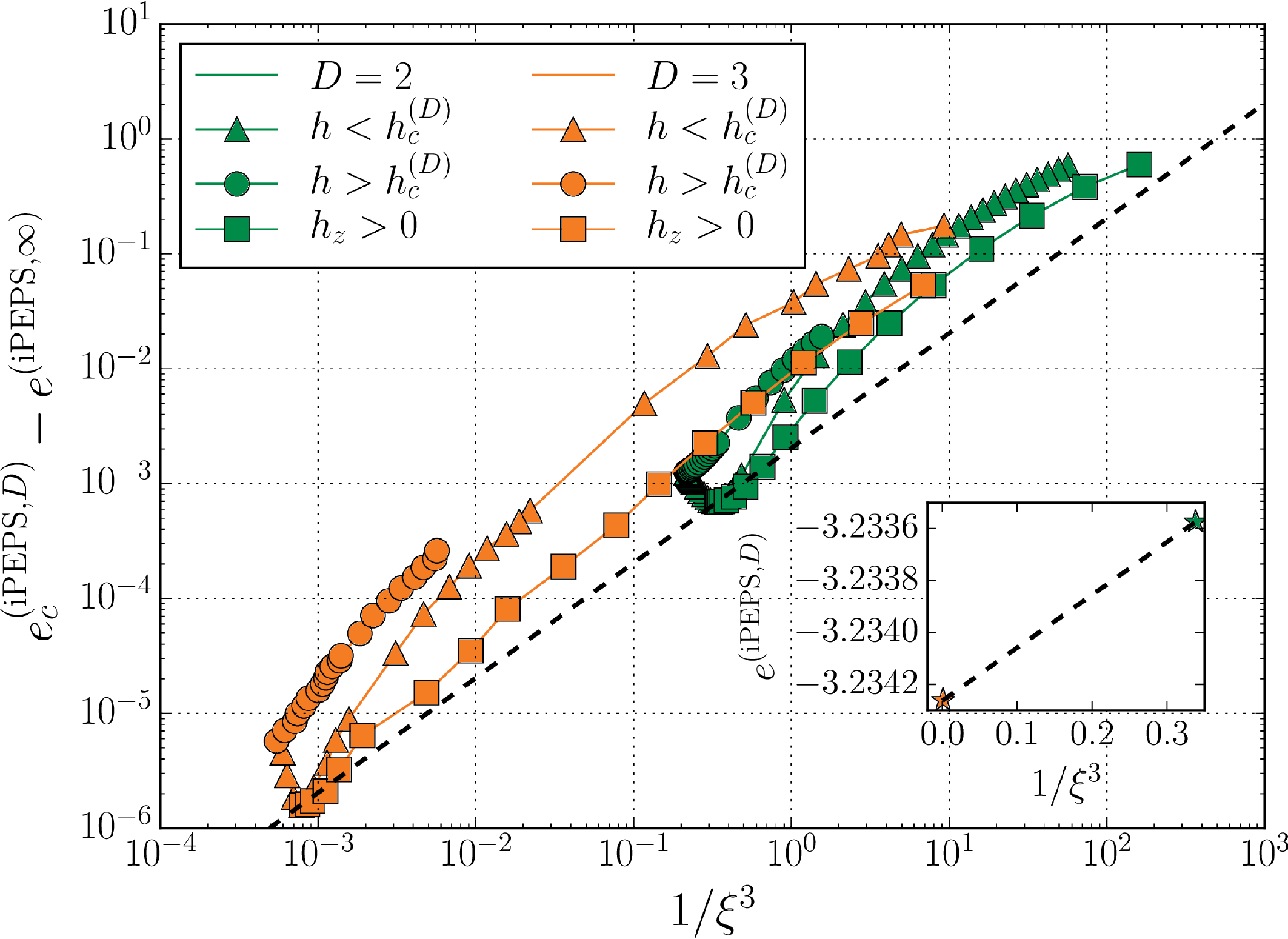}
 \caption{\label{fig:iPEPS_TFI_Casimir} Variational excess critical energy density $\Delta e_c=e_{c}^{(\mathrm{iPEPS},D)}-e_{c}^{(\mathrm{iPEPS},\infty)}$ 
 of transverse field Ising iPEPS wave functions optimized for various $(h,h_z)$ as a function of their $1/\xi^3$. The variational excess energy density corroborates
 the advocated $1/\xi^3$ scaling. Inset: Fit according to Eq.~\eqref{eq:ipeps_casimir}, yielding $e_{c}^{(\mathrm{iPEPS},\infty)}$ and $\alpha^\mathrm{(3d\ Ising\ CFT)}_\mathrm{iPEPS}$
 in \eqref{eq:ipeps_casimir_value}.}
 \end{figure}

\subsubsection{Critical energy convergence}

We now investigate the energy convergence of the TFI model at its critical point $h=h_c,h_z=0$ 
for increasing bond dimension $D$. It is one of the open problems in practical iPEPS calculations to understand the
convergence of energies as a function of $D$. Here we advocate that the variational energy of an 
optimized iPEPS tensor at the critical point $h=h_x,h_z=0$ can be understood as a particular type of a 
Casimir effect controlled by the correlation length $\xi$.

It is well known that the ground state energy density $e=E/N_\mathrm{sites}$ of a 3d quantum critical system
in a torus geometry with modular parameter $\bm{\tau}$~\cite{Schuler2016} is given as
\begin{equation}
e(L)=e(\infty)-\frac{\alpha^\mathrm{QCP}_{\bm{\tau}}\times v}{L^3}, 
\end{equation}
where $L$ denotes the linear extent of the torus, $v$ is the \dquote{speed of light}, e.g.~the critical spin wave 
velocity in a TFI model, and $\alpha^\mathrm{QCP}_{\bm{\tau}}$ is a $\bm{\tau}$ dependent 
Casimir amplitude which otherwise depends only on the universality class of the quantum critical point (QCP),
for example for the 3d Ising CFT and a square torus with periodic boundary conditions $\alpha^\mathrm{(3d\ Ising\ CFT)}_{\bm{\tau}=\bm{i}}=+0.35(2)$
according to Ref.~\cite{Hamer2000}.

We now postulate that our iPEPS setup can be considered as a new, distinct geometry with its own
Casimir amplitude $\alpha^\mathrm{(3d\ Ising\ CFT)}_\mathrm{iPEPS}$, where however the length of the torus
is replaced by the correlation length $\xi$, such that
\begin{equation}
\label{eq:ipeps_casimir}
e(\xi)=e(\infty)-\frac{\alpha^\mathrm{(3d\ Ising\ CFT)}_\mathrm{iPEPS}\times v}{\xi^3}.
\end{equation}
We stress that this Ansatz is in agreement with the expected scaling behavior of the one-point function of the 
stress-energy tensor $T$, whose scaling dimension in $d=3$ is $\Delta_T=3$.
Since we only have two values of $D$, it is a priori hard to determine the validity of the postulated
energy convergence form. Let us nevertheless use the best variational energies at $h=h_c,h_z=0$
for $D=2$ and $D=3$, together with the literature value of $v=3.323(33)$~\cite{Schuler2016} to estimate
\begin{equation}
\label{eq:ipeps_casimir_value}
\alpha^\mathrm{(3d\ Ising\ CFT)}_\mathrm{iPEPS}\approx -0.00061,\quad e(\infty)\approx-3.2342623
\text{.}
\end{equation}
If correct, the scaling hypothesis~Eq.~\eqref{eq:ipeps_casimir} combined with the very small iPEPS Casimir amplitude 
would explain the spectacular accuracy of the $D=3$ results. The $D=3$ correlation length of beyond 10 lattice sites gives
an energy correction proportional to $1/\xi^3\lesssim 10^{-3}$, while the Casimir amplitude 
$\alpha^\mathrm{(3d\ Ising\ CFT)}_\mathrm{iPEPS}$ 
is itself three orders of magnitude smaller than the square torus amplitude. So multiplying these two factors we are led to 
conjecture that the extrapolated iPEPS energies are accurate to about $10^{-6}$ already at $D=3$! This result might  
explain why $D=4$ simulations are so demanding, as the expected remaining energy gains are tiny and are accompagnied 
with wave functions bestowed with large correlation lengths, which are even harder to contract accurately.

In order to corroborate the advocated scaling Ansatz in Eq.~\eqref{eq:ipeps_casimir} we plot the {\em critical} (i.e.~evaluated at $h=h_c,h_z=0$) 
variational energies $e_{c}^{(\mathrm{iPEPS},D)}$ of all 
available iPEPS wave functions in a common plot, c.f.~Fig.~\ref{fig:iPEPS_TFI_Casimir}. In the inset we show the fit according to
Eq.~\eqref{eq:ipeps_casimir} of the two data points with the best $D=2$ and $D=3$ critical energies. The data in the main plot is seen
to approximately scale according to $ \Delta e_c \propto 1/\xi^3$ over several orders of magnitude, thus providing nontrivial empirical evidence in favor of the validity
of the Ansatz~\eqref{eq:ipeps_casimir}~\footnote{We actually believe that the data in Fig.~\ref{fig:iPEPS_TFI_Casimir} ultimately organizes in several
families with different Casimir amplitudes, but all sharing a $1/\xi^3$ scaling.}.

\section{Continuous symmetry breaking}
\label{sec:symmbreaking}

\begin{figure*}
 \includegraphics[width=0.45\linewidth]{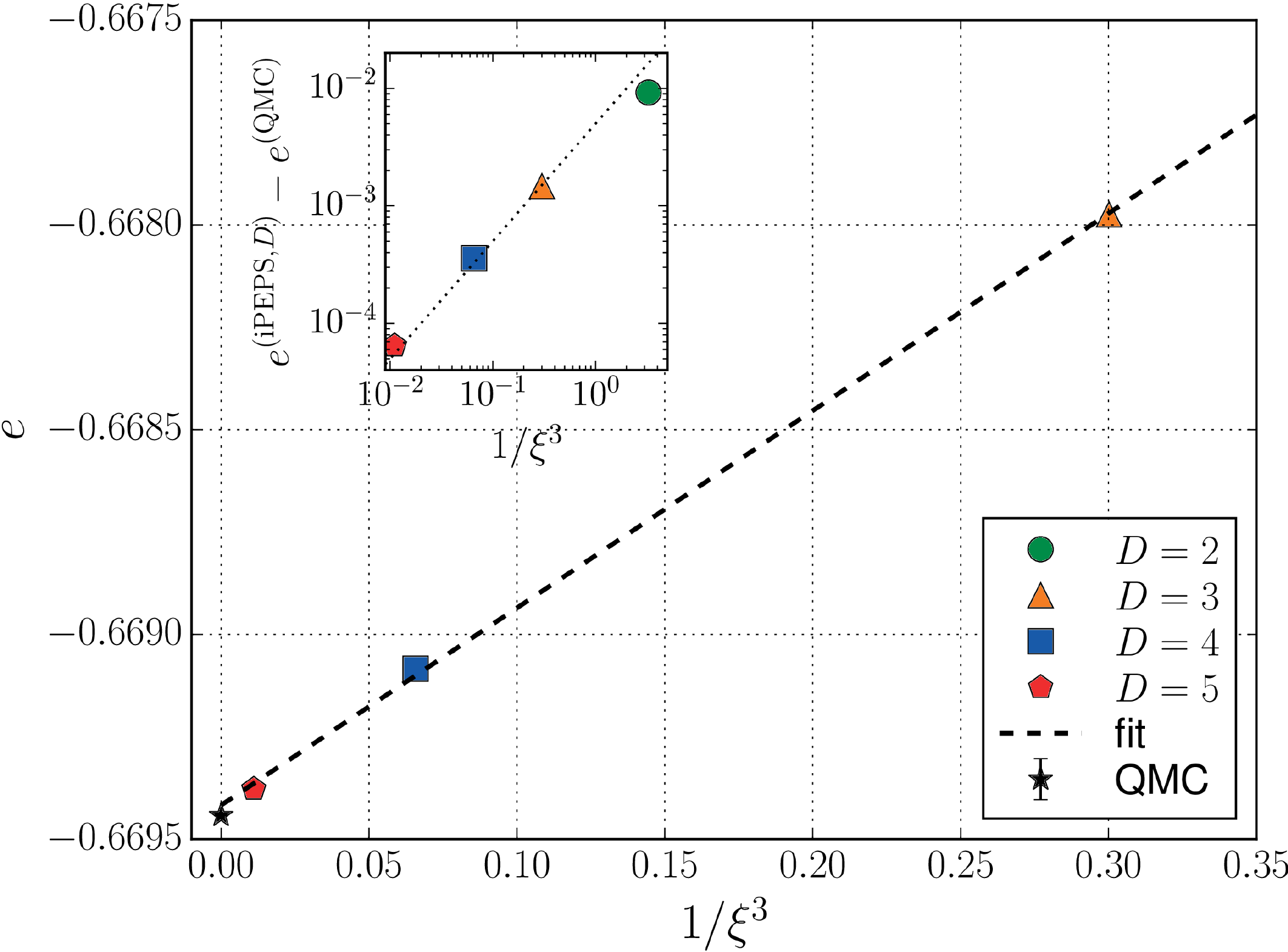}
 \hspace{5mm}
 \includegraphics[width=0.45\linewidth]{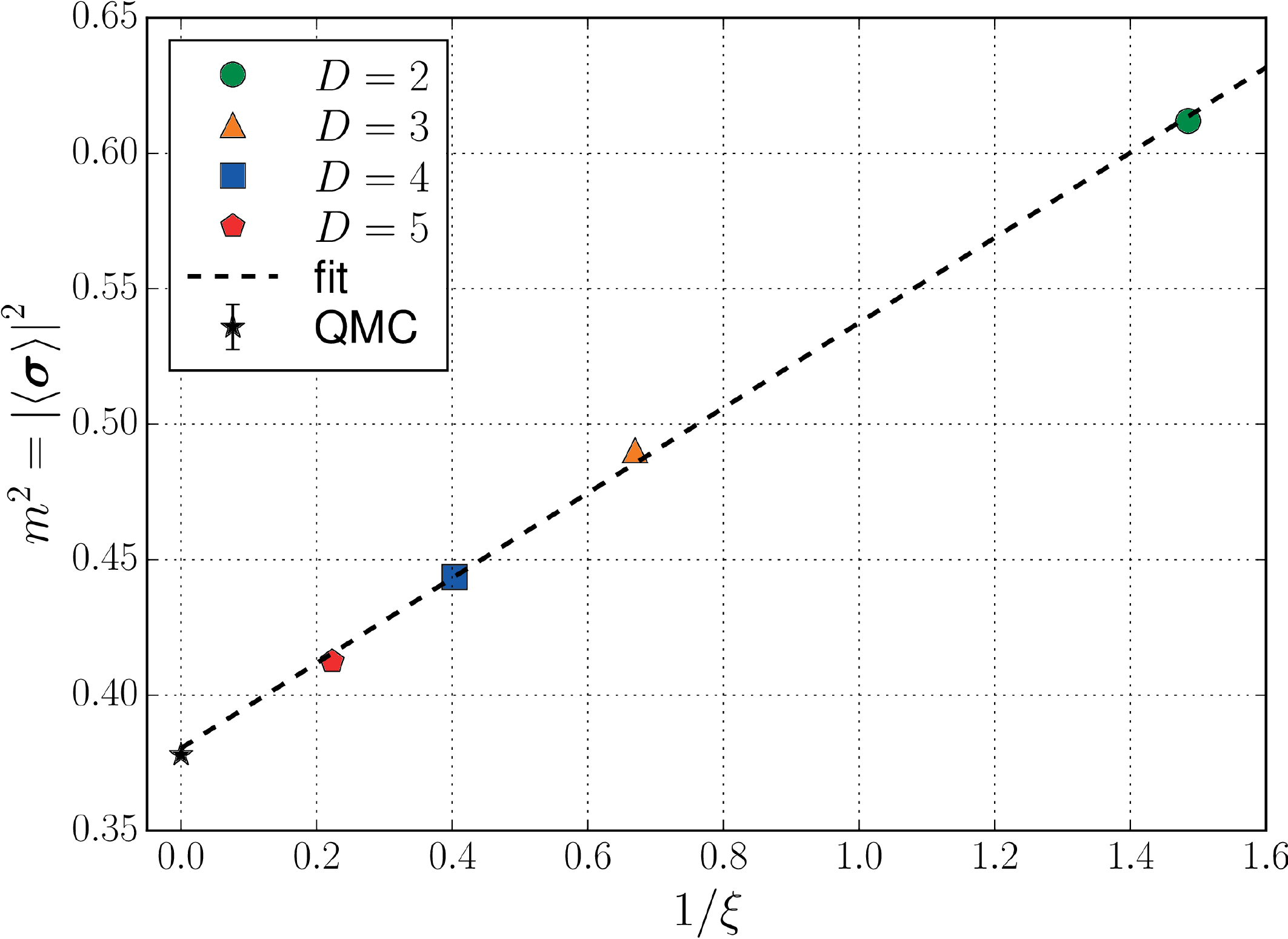}
 \caption{\label{fig:iPEPS_HB} $S=1/2$ Heisenberg antiferromagnet: iPEPS data for the ground state energy per site $e$ (left panel) and the order parameter squared $m^2$ (right panel). 
 We plot the data as a function of the expected $1/\xi^3$ (for $e$) and $1/\xi$ (for $m^2$) dependence. The linear fits to the $D=3,4,5$ results
 in the left and all $D\geq 2$ in the right panel extrapolate closely to the high precision Quantum Monte Carlo reference results. The inset in the left
 panel highlights the overall $1/\xi^3$ convergence of the energy per site.
 }
\end{figure*}

In this section we want to explore the properties of iPEPS wave functions as they 
represent or approximate quantum many body states which exhibit continuous symmetry breaking
in (2+1) space-time dimensions. This is a rather ubiquitous phenomenon, ranging from magnetic 
order in $O(N)$ symmetric quantum magnets with $N\geq2$, to bosonic and fermionic superfluids, to superconductors 
and Goldstone phases in high-energy physics. In these systems the continuous symmetry is spontaneously broken, 
accompanied by the the appearance of a finite order parameter $|\mathbf{M}|>0$. Another hallmark is the required 
presence of gapless excitations (so called Goldstone modes), which are the soft long-wavelength modes of order parameter variations.
We will study the $S=1/2$ Heisenberg antiferromagnet and the $S=1/2$ XY model, both on the square lattice, as paradigmatic
examples for $O(3)$ and $O(2)$ continuous symmetry breaking with a three- respectively two-component vector order parameter.

\subsection{Overview}
The field theoretical description of collinear magnetic order in $O(N)$ quantum magnets relies often on a quantum non-linear sigma model (NLSM) formulation, or on a description in terms of the symmetry breaking phase in an $N$-component interacting $\phi^4$ theory. The gapless
Goldstone modes are also known as spin waves in the magnetic context and taken in isolation they behave as a collection of free 
massless scalar fields. These gapless modes are responsible e.g.~for the algebraic decay of spin-spin correlations to their limiting value
and the finite size corrections of the energy or the order parameter. These linearly dispersing modes require a (2+1)d Lorentz-invariant 
description at low energies, similar to the Ising CFT discussed before.

The quantum non-linear $O(N)$ sigma model is described in detail in Ref.~\cite{Chakravarty1989}. For our purpose it is sufficient to know
that this is basically a hydrodynamic theory 
of quantum magnets with collinear order.
Its description, being hydrodynamic, relies only on a handful
of effective parameters entering the description: the spin stiffness, $\rho_s$, the spin wave velocity $v$, the transverse susceptibility, $\chi_\perp$,
and the squared order parameter in the thermodynamic limit, $m^2(\infty)$. The first three parameters are actually related via $v^2=\rho_s/\chi_\perp$.
Similar to the finite size corrections to the ground state energy discussed in the quantum critical context, the finite size corrections to the ground state
energy and the order parameter have been derived for the quantum non-linear $O(N)$ sigma model 
in Refs.~\cite{Neuberger1989,Nelson1989,Hasenfratz1990,Hasenfratz1993,Sandvik1997}.
The finite size corrections to the ground state energy $e=E/N_\mathrm{sites}$ in $d=3$ are as follows:
\begin{eqnarray}
e(L)&=&e(\infty)- \left[ \alpha^\mathrm{NLSM}_\mathrm{shape/bc} \left(\frac{N-1}{2}\right) v \right] \frac{1}{L^3} \nonumber\\
&&+\frac{(N-1)(N-2)}{8}\frac{v^2}{\rho_s L^4}+\mathcal{O}\left(\frac{1}{L^5}\right)\ .
\end{eqnarray}
For a square torus with periodic boundary conditions $\alpha^\mathrm{NLSM}_\mathrm{\mathbf{\tau}=\mathbf{i}}\approx1.437745$ has been
obtained.
The finite size correction for the magnetic order parameter squared are as follows:
\begin{equation}
\frac{m^2(L)}{m^2(\infty)}=1+ \left[ \mu^\mathrm{NLSM}_\mathrm{shape/bc} \left(\frac{N-1}{2}\right) \frac{v}{\rho_s}\right] \frac{1}{L} +\mathcal{O}\left(\frac{1}{L^2}\right)
\text{.}
\end{equation}
For a square torus with periodic boundary conditions $\mu^\mathrm{NLSM}_\mathrm{\mathbf{\tau}=\mathbf{i}}\approx 0.62075$ has been
found.

As we will see below, our variationally optimized iPEPS wave functions have a finite correlation length $\xi$, which depends on the 
model and on the bond dimension $D$. We now conjecture the following $d=3$, finite $\xi$ corrections for the ground state energy density $e$ and
the magnetic order parameter squared $m^2$,
\begin{equation}
e(\xi) = e(\infty)- \left[\alpha^\mathrm{NLSM}_\mathrm{iPEPS} \left(\frac{N-1}{2}\right) v\right] \frac{1}{\xi^3} 
+ \mathcal{O}\left(\frac{1}{\xi^4}\right)\ , \label{eqn:e_xi}
\end{equation}
and
\begin{equation}
\frac{m^2(\xi)}{m^2(\infty)}= 1+ \left[\mu^\mathrm{NLSM}_\mathrm{iPEPS} \left(\frac{N-1}{2}\right) \frac{v}{\rho_s}\right] \frac{1}{\xi} + \mathcal{O}\left(\frac{1}{\xi^2}\right)\ . \label{eqn:m2_xi}
\end{equation}
The potential power of these formulas lies in the fact that one can extrapolate the results at finite $\xi(D)$ to the limit $\xi\rightarrow \infty$ based on iPEPS data fits
to the above formulas. Furthermore it is possible to predict the finite size corrections for other microscopic models, once the \dquote{universal} values of 
$\alpha^\mathrm{NLSM}_\mathrm{iPEPS}$ and $\mu^\mathrm{NLSM}_\mathrm{iPEPS}$ are determined.  The knowledge of $N$, $v$ and $v/\rho_s$
allows then to quantitatively predict the slope of the finite size corrections. Conversely it might become possible to estimate $v$ and $v/\rho_s$ based on precise 
iPEPS data for a model with a known value for $N$.

\begin{figure*}
 \includegraphics[width=0.45\linewidth]{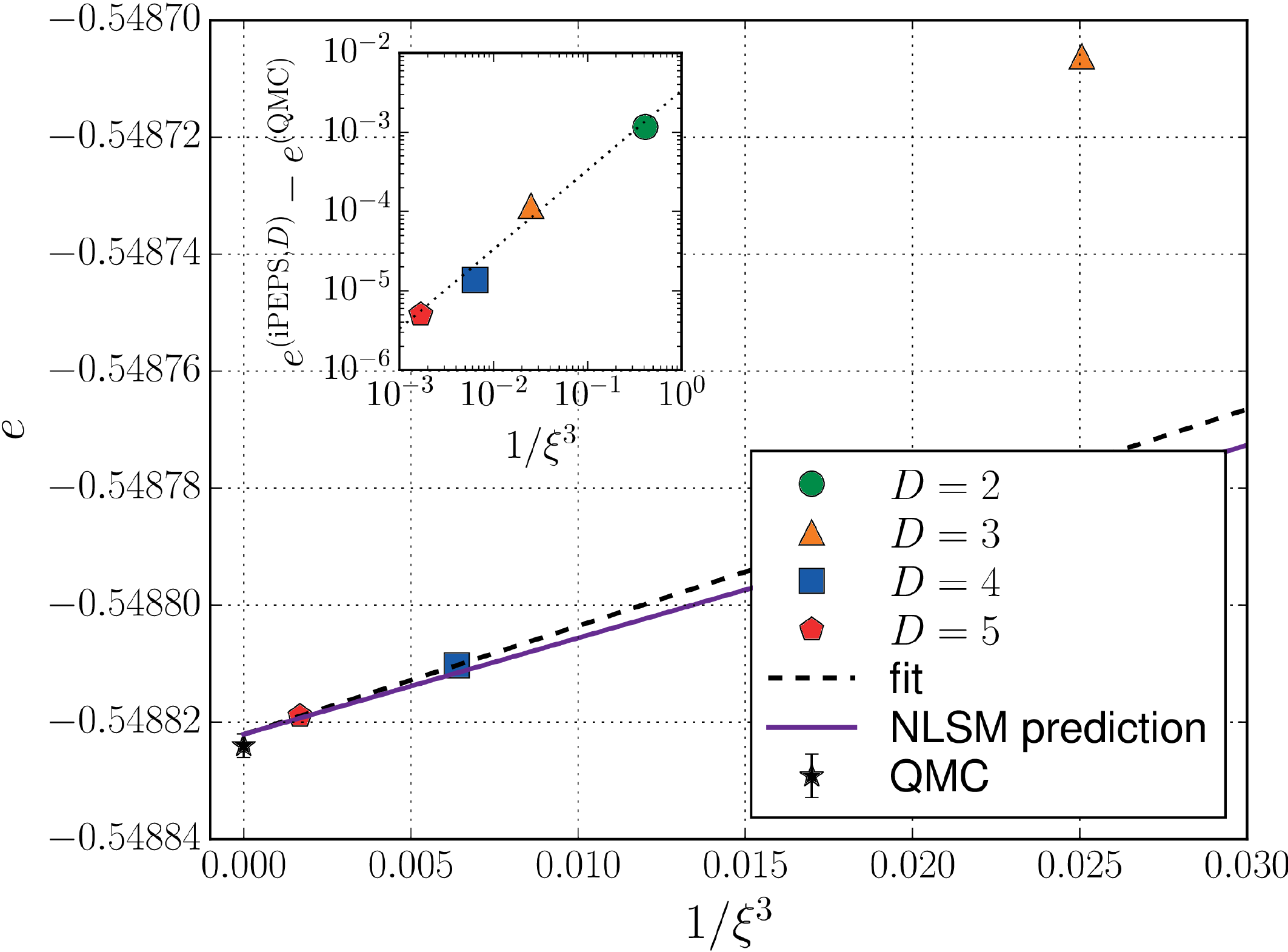}
 \hspace{5mm}
 \includegraphics[width=0.45\linewidth]{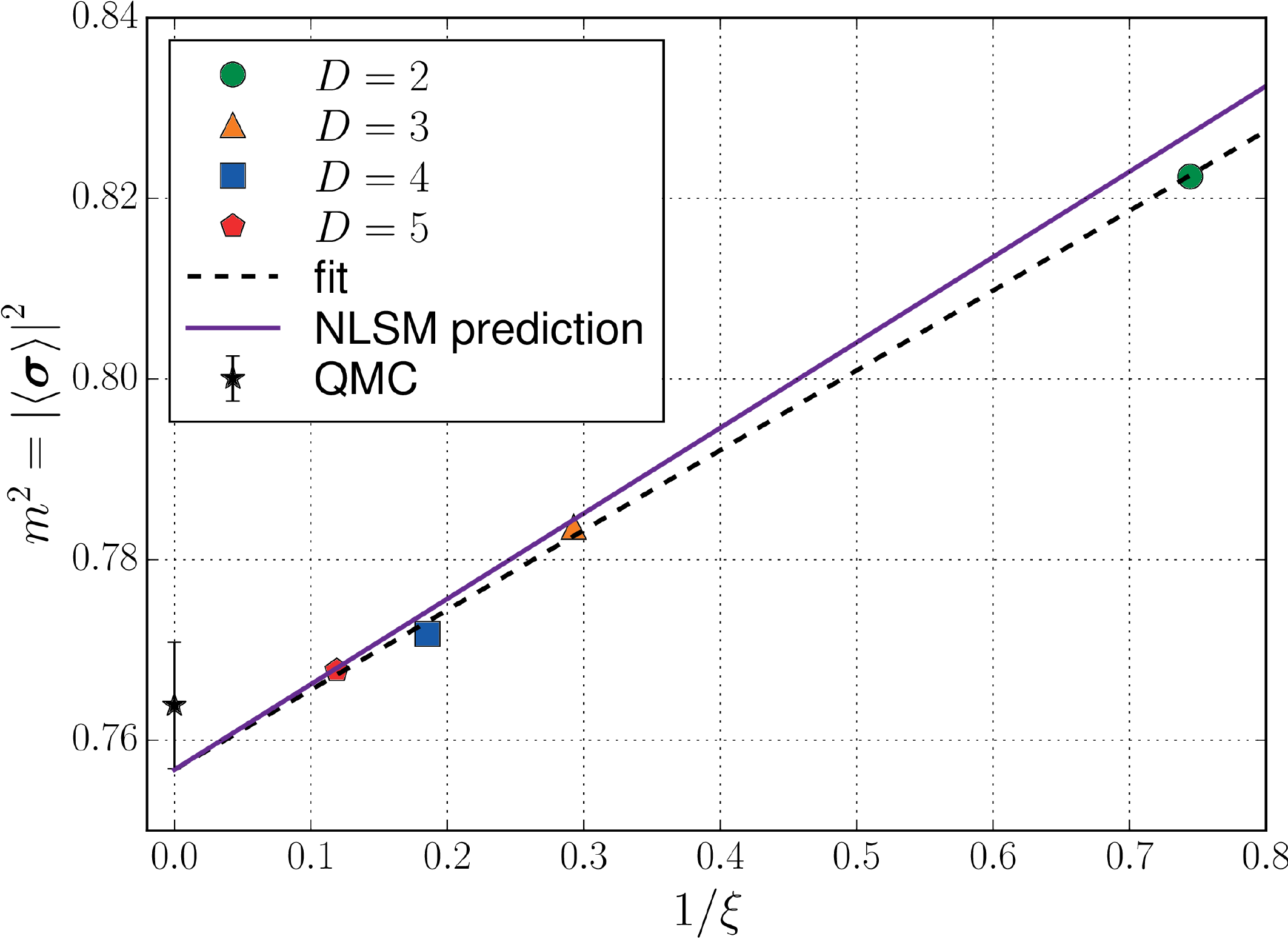}
 \caption{\label{fig:iPEPS_XY}
 $S=1/2$ XY ferromagnet: iPEPS data for the ground state energy per site $e$ (left panel) and the order parameter squared $m^2$ (right panel). 
 We plot the data as a function of the expected $1/\xi^3$ (for $e$) and $1/\xi$ (for $m^2$) dependence. The linear fits to the $D=4,5$ results
 in the left and all $D\geq 2$ in the right panel extrapolate reasonably closely to the Quantum Monte Carlo results, and may well be more accurate
 than the somewhat antiquated QMC results. The inset in the left
 panel highlights the overall $1/\xi^3$ convergence of the energy per site. In both panels we include a prediction based on the conjectured finite
 $\xi$ formulas for the non-linear sigma model (NLSM), c.f.~Eqs.~\eqref{eqn:e_xi}, \eqref{eqn:m2_xi} and main text.
 }
\end{figure*}

\subsection{$S=1/2$ antiferromagnetic Heisenberg model}

The $S=1/2$ antiferromagnetic Heisenberg model has also been studied frequently using
finite size PEPS and iPEPS approaches in the 
past~\cite{Gu2008,Bauer2009,Bauer2011,Lubasch2014,Corboz2016a,Corboz2016b,Vanderstraeten2016}.
The ground state of this model has antiferromagnetic N\'eel order, which breaks the continuous $O(N{=}3)$ 
rotation symmetry spontaneously down to a residual $O(2)$ symmetry. The presence of $N-1=2$ Goldstone 
modes leads to an algebraic decay of the two-spin correlation function. 

The Hamiltonian is defined as
\begin{equation}
H_\mathrm{HB}=J \sum_{\langle i,j\rangle} \left( S^x_iS^x_j + S^y_iS^y_j + S^z_iS^z_j\right) ,
\end{equation}
with $J=1$ and the $S^\alpha_i$ are spin $1/2$ operators. In order to be able to work with a 
single-site unit cell we perform a spin rotation on one N\'eel sublattice, which negates 
the sign of the $y$ and $z$ parts of the interactions in the actual calculations.

We proceed optimizing the variational energies of iPEPS tensors for $D=2,3,4,5$. Then we measure 
the order parameter squared $m^2=|\langle \bm{\sigma}_i\rangle|^2$ (i.e.~the maximum possible for the order parameter 
squared amounts to one), as well as the correlation length $\xi$. These correlation lengths are finite and range from
$\xi(D=2) \approx 0.7$ to $\xi(D=5) \approx 4.5$.
The correlation lengths are thus substantially smaller than those
of the critical transverse field Ising model at $D=3$.

In Fig.~\ref{fig:iPEPS_HB} we present the energy per site $e$ as function of $1/\xi^3$ in the left panel and $m^2$ as a function
of $1/\xi$ in the right panel. This is the conjectured $\xi$ scaling form of Eqs.~\eqref{eqn:e_xi} and \eqref{eqn:m2_xi}. 
It is striking that for both observables a linear fit leads to accurate extrapolations to the limit 
$\xi\rightarrow\infty$, when compared to high precision QMC reference values~\cite{Sandvik1997,Sandvik2010}. We fit
the largest three $D$ values for the energy $e$ and all the $D$ values for $m^2$ and obtain the following iPEPS $\xi\rightarrow \infty$ 
estimates: $e_\mathrm{HB}\approx-0.669417$ and $m^2_\mathrm{HB}\approx 0.380$. Using the 
iPEPS fit slopes, the value $N=3$ and the known QMC values of the hydrodynamic parameters $v$ and $v/\rho_s$~\cite{Sandvik1997,Jiang2011}, 
we can then proceed to determine 
\begin{equation}
\alpha^\mathrm{NLSM}_\mathrm{iPEPS}\approx-0.0029
\label{eqn:alpha_NLSM}
\end{equation} 
and 
\begin{equation}
\mu^\mathrm{NLSM}_\mathrm{iPEPS}\approx+0.045
\label{eqn:mu_NLSM}
\text{.}
\end{equation} 
Note that similar to the Ising Casimir amplitude, the NLSM energy Casimir amplitude is three orders of
magnitude smaller than the square torus one, highlighting that an iPEPS calculation at a certain $\xi$ should be considered three orders of magnitude more
accurate than a square torus with $L\sim\xi$. The iPEPS order parameter amplitude is however only one order of magnitude smaller than
the square torus result.

\subsection{$S=1/2$ XY model}

The $S=1/2$ XY model has also been investigated with PEPS and iPEPS in the 
past~\cite{Jordan2009,Bauer2009,Vanderstraeten2016}. The ground state of this model has ferromagnetic 
order in the $x{-}y$ spin plane, which breaks the continuous $O(N{=}2)$ in-plane rotation symmetry spontaneously 
down to a residual discrete $\mathbb{Z}_2$ symmetry. The presence of $N-1=1$ Goldstone mode leads to an algebraic 
decay of the two-spin correlation function. 

The Hamiltonian is defined as,
\begin{equation}
H_\mathrm{XY}=-J \sum_{\langle i,j\rangle}  \left( S^x_i S^x_j + S^y_i S^y_j \right) ,
\end{equation}
with $J=1$. Note that for this model the two choices of the sign of $J$ can be
mapped into each other. Since we want to use a single site iPEPS unit cell we adopt the
ferromagnetic sign convention.

We proceed optimizing the variational energies of iPEPS tensors for $D=2,3,4,5$. Then we measure 
the order parameter squared $m^2=|\langle \bm{\sigma}_i\rangle|^2$ (i.e.~the maximum possible for the order parameter 
squared amounts to one), as well as the correlation length $\xi$. These correlation lengths are again finite and range from
$\xi(D=2) \approx 1.3$ to $\xi(D=5) \approx 8.4$.
The correlation lengths for each $D$ are roughly a factor two 
larger than for the Heisenberg model. So it seems that the smaller number of Goldstone modes has a beneficial effect 
on the growth of the correlation lengths.

In Fig.~\ref{fig:iPEPS_XY} we present the energy per site $e$ as function of $1/\xi^3$ in the left panel and $m^2$ as a function
of $1/\xi$ in the right panel. This is the conjectured $\xi$ scaling form of Eqs.~\eqref{eqn:e_xi} and \eqref{eqn:m2_xi}. We then fit
the largest two $D$ values for the energy $e$ and all the $D$ values for $m^2$ and obtain the following iPEPS $\xi\rightarrow \infty$ 
estimates: $e_\mathrm{XY}\approx-0.548822$ and $m^2_\mathrm{XY}\approx 0.757$. These values compare well with the QMC 
results of Ref.~\cite{Sandvik1999}. Since the QMC results are relatively antiquated, it is not inconceivable that the iPEPS results are actually
more precise than the QMC results.

We are now also in a position to corroborate the $O(N)$ universality of the conjectured NLSM finite $\xi$ corrections.  
Using the hydrodynamic parameters $v$ and $v/\rho_s$ from Ref.~\cite{Sandvik1999}, the iPEPS amplitudes from Eqs.~\eqref{eqn:alpha_NLSM} and
\eqref{eqn:mu_NLSM} and inserting $N=2$ we arrive at the NLSM predictions, which are shown by solid lines in both panels of Fig.~\ref{fig:iPEPS_XY}. The nice agreement between
the linear fits and the NLSM predictions for the slopes provide further support for the validity and therefore power of the field theoretically inspired finite-$\xi$ correction formulae.

\begin{figure}
  \includegraphics[width=0.95\linewidth]{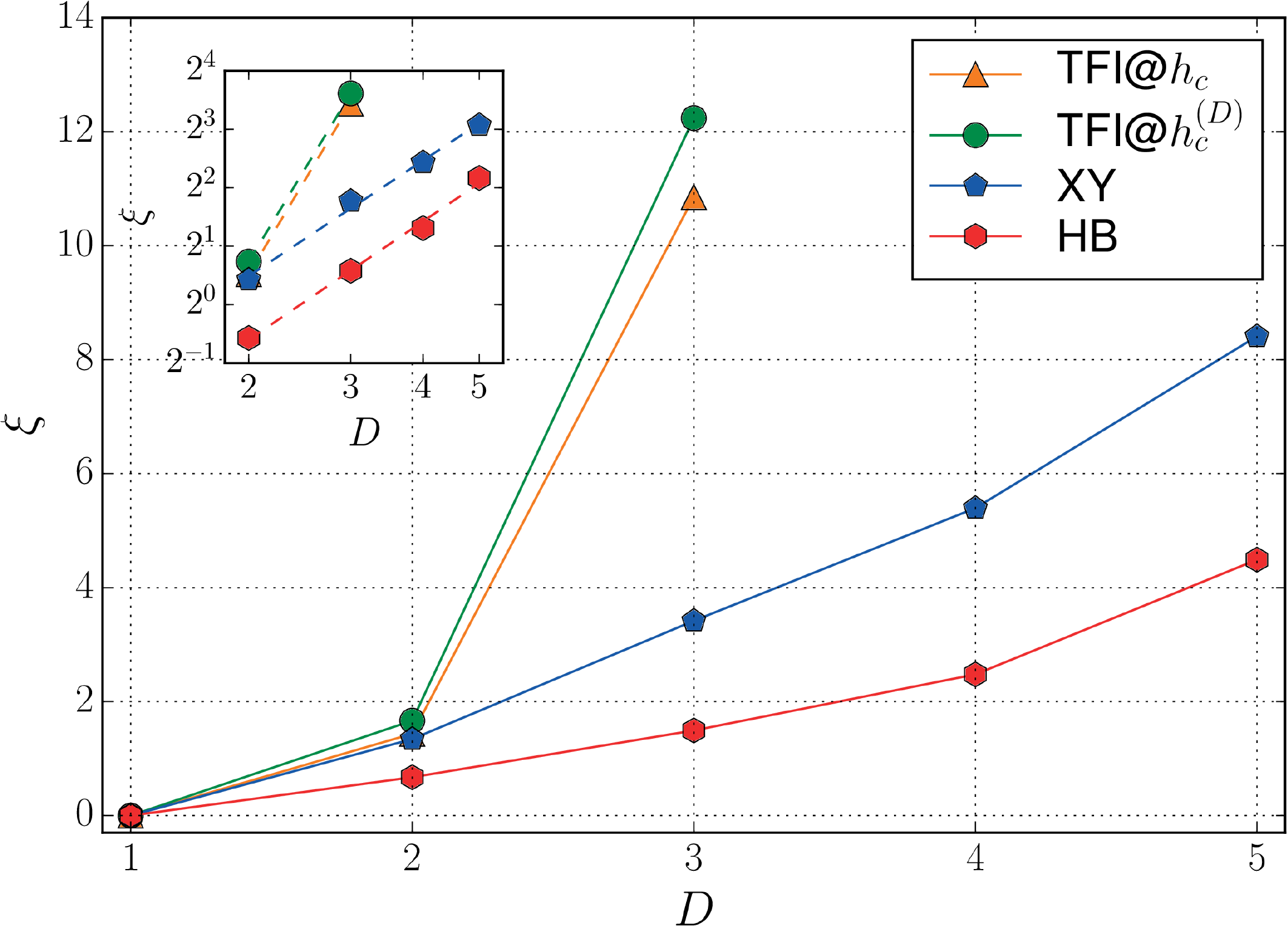}
 \caption{\label{fig:overview_corr_lengths} 
 Overview of the finite $D$ iPEPS correlation lengths $\xi$ observed in the critical transverse field Ising model (TFI) and 
 the continuous symmetry breaking XY and Heisenberg models. The inset displays the same data on logarithmic 
 scales.}
\end{figure}

\section{Discussion and Interpretation}
\label{sec:discussion}

After having studied the three different models we are confronted with the fact
that in all cases the correlation length $\xi(D)$ was finite. While we have developed
a powerful finite $\xi$ scaling framework, where many observables, including energies 
and order parameters, can be analyzed and extrapolated to $\xi\rightarrow\infty$, we are still
left in the dark both regarding the underlying origin of the finite $\xi(D)$ in the first place, and 
regarding the functional dependence of $\xi$ on $D$ for a given model or universality class. 

While we don't yet have compelling answers to both questions so far, we can at least try to 
shed as much light as possible based on our numerical data. In Fig.~\ref{fig:overview_corr_lengths}
we have assembled the correlation lengths observed in the critical transverse field Ising model, which
is described by one of the simplest non-trivial $3d$ CFTs, together with two instances of $O(N)$ 
continuous symmetry breaking phenomena with $N=3$ for the $S=1/2$ antiferromagnetic Heisenberg
model and $N=2$ for the $S=1/2$ XY model. In the inset we also show the $D\geq2$ data in a $\log$-$\log$
plot, yielding some rough estimates for a putative power-law relation $\xi(D)\sim D^\kappa$ (it is not clear that
such a law holds). With the two points for the critical transverse field Ising model we obtain 
$\kappa_\mathrm{(3d\ Ising\ CFT)}\approx 5$, while both continuous symmetry breaking cases seem to share the
same $\kappa_\mathrm{NLSM}\approx 2$. The latter two cases however differ by a factor two in the prefactor,
which incidentally is also the inverse ratio of the number of Goldstone modes. This could mean 
that the XY model has twice as large a correlation length as the Heisenberg model because it only 
has half the number of Goldstone modes. It would be interesting to explore whether the known additive 
logarithmic correction to the entanglement area law in continuous symmetry breaking states~\cite{Metlitski2011}
might be at the origin of this behavior. There the prefactor of the $\log$ contribution is proportional to the 
number of Goldstone modes. This would also explain why the speculative values of $\kappa$ are so different
between the $3d$ Ising CFT and the symmetry breaking cases. In terms of their low-energy degrees of freedom,
the $3d$ Ising CFT and the XY model both contain a single real scalar field each -- an interacting field in the Ising CFT
case and a massless free field in the XY case. One could thus have expected the values of $\kappa$ to be roughly
similar. The crucial difference therefore seems to come from the broken continuous symmetry in the XY model, which is
absent in the Ising CFT.

\begin{figure}[t]
 \includegraphics[width=0.9\linewidth]{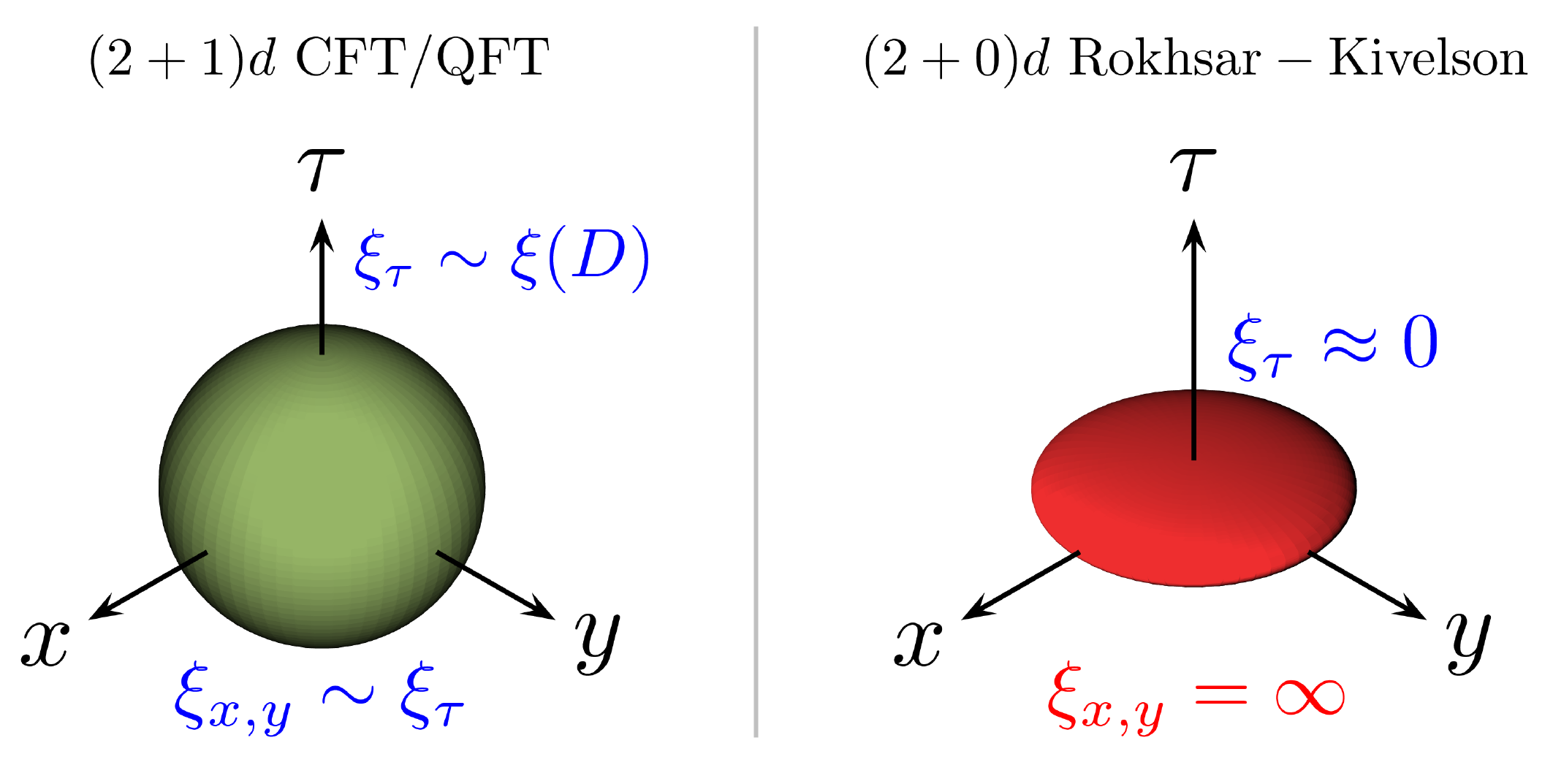}
 \caption{\label{fig:iPEPS_spacetime}
 Cartoon illustration of a the space-time volume sampled in a $(2+1)d$
 iPEPS versus the $(2+0)d$ Rokhsar-Kivelson type iPEPS wavefunctions.
 }
\end{figure}

\begin{figure*}
 \includegraphics[width=0.4\linewidth]{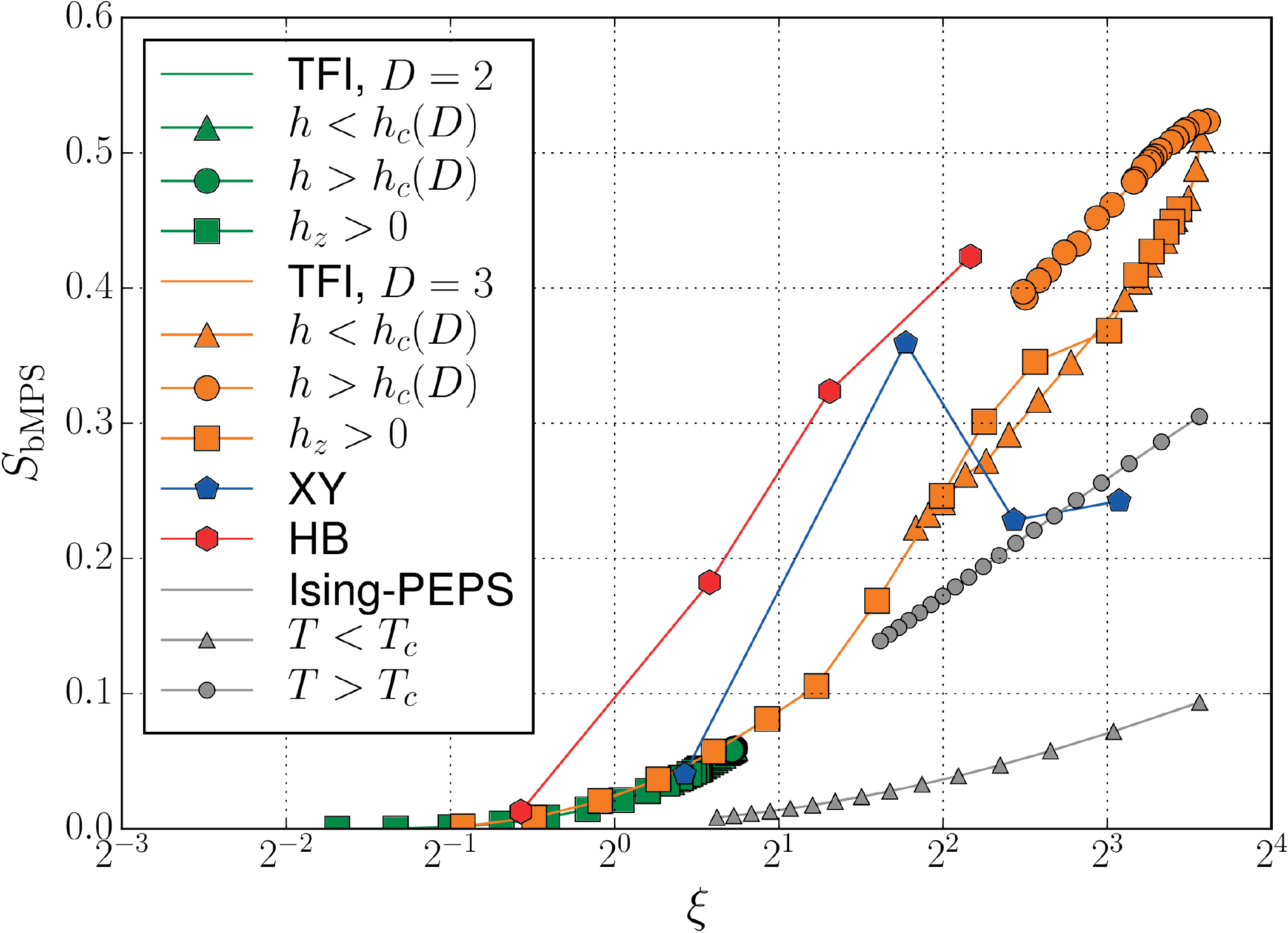}
  \hspace{5mm}
  \includegraphics[width=0.4\linewidth]{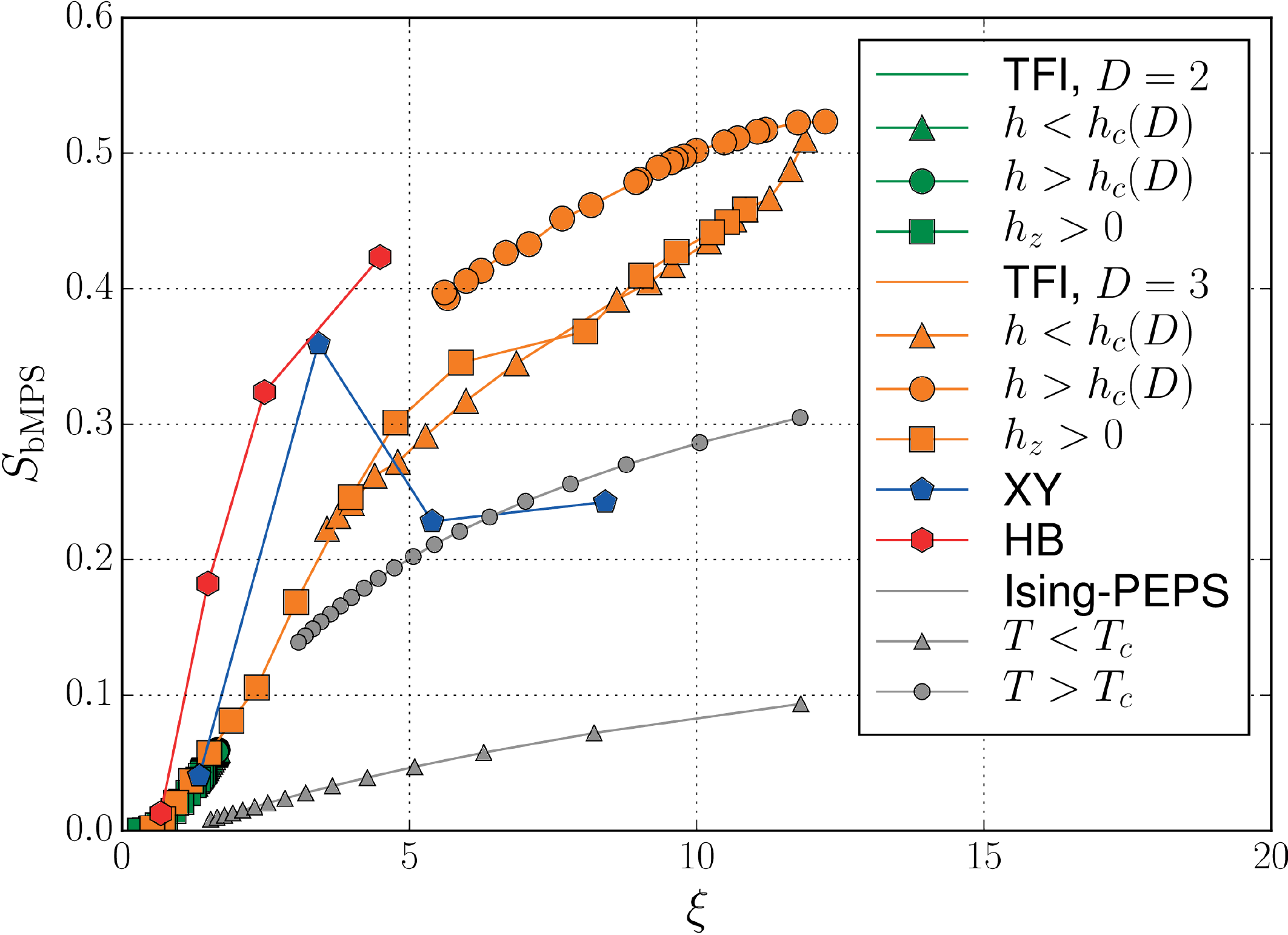}
 \caption{\label{fig:iPEPS_bMPS_S}
 Half chain von Neumann entropy $S_\mathrm{bMPS}$ of the boundary MPS resulting from the contraction of the iPEPS tensors, plotted as a function of $\log \xi$ (left panel) 
 and $\xi$ (right panel). We display
 all available data sets. The transverse field Ising model data is organized according to $D$ and the parameter cut, while the
 Heisenberg and XY model connect all available $D$ values. For comparison we also show entanglement entropies
 for the $(2+0)d$ Ising PEPS evaluated at different temperatures.
 }
\end{figure*}

The most pressing question remains however as to why we only find {\em finite} correlation lengths in the variationally optimized
iPEPS wave functions for massless Lorentz-invariant $(2+1)d$ scenarios. We believe that our observations are actually 
the generic result, and that the previously known examples of iPEPS states with algebraic correlations are fine-tuned and non-generic. As shown
as a cartoon in the left panel of Fig.~\ref{fig:iPEPS_spacetime}, we think of our $(2+1)d$ iPEPS states as wave functions whose correlation functions 
are represented by a path integral with a finite $\xi_\tau(D)$ extent in the (real or imaginary) time direction. The Lorentz-invariance (or Euclidian invariance after a 
Wick-rotation) of the fixed point we try to approximate then forces the spatial correlation lengths to be finite as well. The well known 
iPEPS states with algebraic spatial correlations at finite $D$ can actually be represented by a purely in-plane path integral, where the temporal extent $\xi_\tau$ is basically
zero (right panel). This is certainly true for the $2d$ classical partition function Ising PEPS~\cite{Verstraete2006}, quantum dimer Rokhsar-Kivelson states~\cite{Castelnovo2005,Schuch2012} and 
certain quantum Lifshitz theories~\cite{Isakov2011,Hsu2013}, where there is a built-in space-time asymmetry. Some further evidence supporting this picture might be obtained 
from the entanglement entropies $S_\mathrm{bMPS}$ of the boundary MPS resulting from the contractions of the iPEPS, as it seems plausible that this entanglement entropy
is amplified if the correlation volume extends into the $\tau$ direction. The data shown in Fig.~\ref{fig:iPEPS_bMPS_S}
can indeed be interpreted that the entanglement entropy grows more rapidly with the correlation length $\xi$ in the genuine $3d$ space-time cases, compared to
the quantitatively well understood logarithmic scaling of the $(2+0)d$ wavefunctions, here exemplified by the Ising PEPS at various temperatures~\footnote{The XY model
shows a somewhat irregular behavior, whose origin we do not understand}. It would be interesting to study these boundary MPS entropies more systematically, as they roughly dictate
how large the boundary $\chi$ value in the contraction schemes has to be.

Coming back to the reason for the finite $\xi_\tau(D)$  in the first place: We speculate that finding the dominant eigenvector 
of a plane-to-plane transfer operator along the temporal direction combined with a projection to a finite $D$
iPEPS leads invariably to a finite correlation length in the temporal direction. This view is also supported by a more formal argument
stating that an entropic area law does not automatically imply an efficient iPEPS representation~\cite{Ge2016}.

To a first approximation we understand the finite-$D$ iPEPS to mimic a ground state wave function of the field 
theoretical fixed point perturbed by a certain amount of the most relevant allowed perturbation given the symmetry constraints imposed on the iPEPS. 
Such states are driven away from the gapless situation, leading to a finite correlation length $\xi(D)$.
In the examples studied here this always corresponds to a perturbation by coupling to the magnetic order parameter. 
Imposing symmetry constraints on the tensor might change the nature of the allowed relevant perturbation, and could also affect 
the values of some of the finite $\xi$ correction amplitudes introduced in this work.

\section{Conclusions}
\label{sec:conclusions}

In this paper we have introduced a powerful finite correlation length scaling framework for the
analysis of finite-$D$ iPEPS which have been variationally optimized for
Lorentz-invariant $(2+1)d$ quantum critical or continuous symmetry breaking Hamiltonians.
This framework is important i) to understand the {\em mild} limitations of the finite-$D$ 
iPEPS manifold in representing Lorentz-invariant, gapless many body quantum states and
ii) to put forward a practical scheme in which the finite correlation length $\xi(D)$ and field theory
inspired formulae can be employed to extrapolate the data to infinite correlation length, i.e. to the
thermodynamic limit. We have shown that some of the amplitudes entering the equations have a 
field theoretical interpretation and are therefore universal to some degree. We have demonstrated 
the power of the method for the energy convergence in all three considered models, and for  order parameter
extrapolations for the continuous symmetry breaking models, where the previously employed $1/D$ extrapolation
schemes performed poorly. We believe that another advantage of this approach is also that the variational
quality of an iPEPS tensor and the resulting correlation length (and other observables) seem to go hand in hand 
in the vicinity of a local optimum, in such a way that different data points still lie on the same finite $\xi$ extrapolation curve. 

We have also carefully analyzed the critical behavior of the transverse field Ising model in the
vicinity of its critical point by measurements of local observables as a function of the two magnetic fields
and we were able to obtain critical exponents within a few percent of the most recent conformal bootstrap 
results already at bond dimension $D=3$.

The finite correlation length scaling framework opens the way for further exploration of quantum matter with
an (expected) Lorentz-invariant, massless low-energy description. Within the quantum critical or CFT related 
questions one could explore the Wilson-Fisher $O(N)$ 
fixed points beyond the $N=1$ case studied here, Gross-Neveu-Yukawa universality classes arising in
interacting Dirac fermion systems, or QED$_3$ related behavior of gapless 
quantum spin liquids or deconfined criticality. In the context of continuous symmetry breaking various superfluid
and superconducting phases in bosonic and fermionic systems should be described by the $O(2)$ symmetry breaking
results established in this work. Further directions are non-collinear magnetic order or $\mathrm{SU}(N)$ continuous symmetry
breaking. Ultimately one should also explore the occurence of finite correlation lengths and their scaling in 
interacting systems with a Fermi surface.

On the fundamental level it will be important to develop an understanding of the $\xi(D)$ relation, that we started to 
explore here. Our preliminary results indicate a $\xi \sim D^\kappa$ scaling with values of $\kappa$ which could be
universal. In the $(1+1)d$ iMPS context, the values of $\kappa$ depend only on the central charge $c$~\cite{Pollmann2009}. It would be 
interesting to understand  in $(2+1)d$ whether $\kappa$ also depends only on some universal data, such as in the $F$-theorem~\cite{Casini2011,Grover2014}.

{\em Note:} Similar results have been reported by P.~Corboz, P.~Czarnik, G.~Kapteijns, and L.~Tagliacozzo, see arXiv:1803.?????

\acknowledgments
We acknowledge discussions with P.~Corboz, J.~Haegeman, G.~Misguich, L.~Tagliacozzo, L. Vanderstraeten, F. Verstraete and G.~Vidal.
We thank P.~Corboz for bringing Ref.~\cite{rams2018precise} to our attention.
We acknowledge support by the Austrian Science Fund FWF within the DK-ALM (W1259-N27) and the SFB FoQus (F-4018)
The computational results presented have been achieved in part using the HPC infrastructure LEO of the University of Innsbruck.
The computational results presented have been achieved in part using the Vienna Scientific Cluster (VSC).

\bibliographystyle{apsrev4-1_custom}
\bibliography{fes,peps} 

\appendix
\renewcommand\thefigure{\thesection.\arabic{figure}}
\renewcommand\thetable{\thesection.\arabic{table}}

\setcounter{figure}{0}
\setcounter{table}{0}

\section{Further results on the transverse field Ising model}
\label{app:mx_h}
\begin{figure*}
  \includegraphics[width=0.45\linewidth]{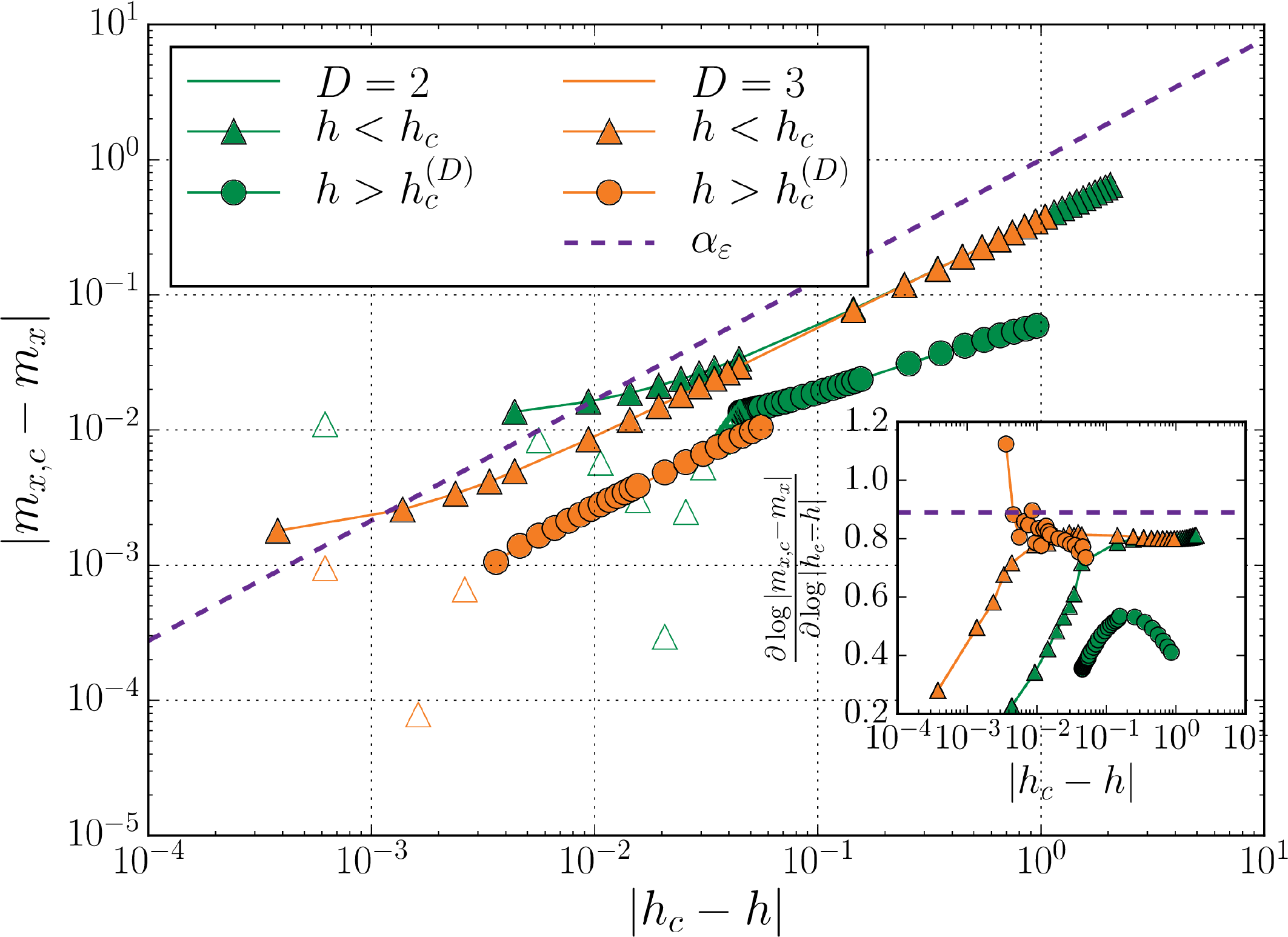}
  \hspace{5mm}
  \includegraphics[width=0.45\linewidth]{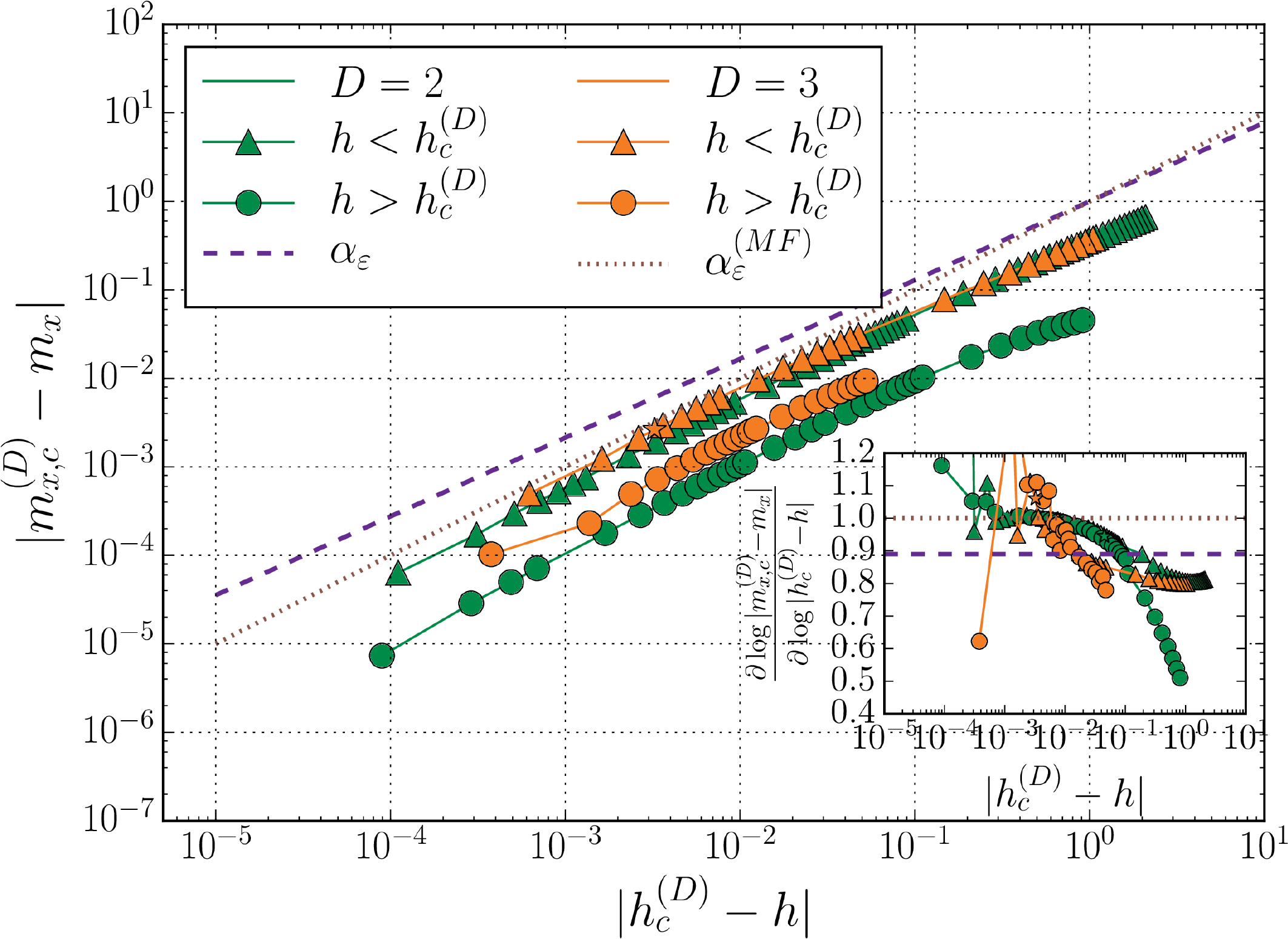}
  \caption{\label{fig:TFI_mx_h} Analysis of the critical behavior of the $m_x$ magnetizations in the transverse field Ising model in the vicinity of the critical point at $h=h_c,h_z=0$ as a function of the transverse field. The left panel shows $|m_{x,c}-m_x|$ as a function of $|h_c-h|$ and the right panel $|m_{x,c}^{(D)}-m_x|$ as a function of $|h_c^{(D)}-h|$. In the insets numerical derivatives are presented (where the hollow symbols of the left panel are excluded as they cause divergences in the derivative).
  The inset of the right panel shows mean-field behavior of the $D=2$ data at small $|h_c^{(D)}-h|$
  while the $D=3$ data is too noisy to allow firm conclusions.
  The expected exponents are tabulated in Tab.~\ref{tab:IsingCFT}.
}
\end{figure*}

In this appendix the critical behavior of the $m_x$ magnetizations as a function of the transverse field $h$ in the transverse field Ising model in the vicinity of the critical point at $h=h_c$, $h_z=0$ is presented.
In the left panel of \figref{fig:TFI_mx_h} we plot $|m_{x,c}-m_x|$ as a function of $ |h_c-h|$ on a $\log$-$\log$ scale together with a straight line showing the expected slope $\alpha_\varepsilon$ as a guide to the eye. In the corresponding inset we numerically calculate the derivative (excluding the hollow symbols as they cause divergences in the derivative).
The data shows convergence towards the expected value for the critical exponent $\alpha_\varepsilon$ from $D=2$ to $D=3$, although the noise on this data is relatively large compared to the data presented in \figref{fig:iPEPS_TFI_criticality}.
This might be due to the uncertainty in the estimation of $m_{x,c}$.

In the right panel of \figref{fig:TFI_mx_h} $|m_{x,c}^{(D)} - m_x|$ is presented as a function of $|h_c^{(D)} - h|$ on a $\log$-$\log$ scale and the inset contains the corresponding numerical derivatives. 
The $D=2$ data clearly shows mean-field behavior at small $|h_c^{(D)}-h|$, but the derivatives for $D=3$ might already start to form an intermediate plateau.
We expect data for ${D>3}$ to give a better understanding of how this plateau is formed.

\section{Variational iPEPS Energies}

For reference we provide in Tab.~\ref{tab:variational_energies} our best variational iPEPS energies for the three models considered
in this paper. The transverse field Ising model is at its critical point $h=h_c,h_z=0$. 

\begin{table}[b]
\begin{tabular}{|c|c|c|}
	\hline
	Model        & $D$ & Energy / $J$ \\
	\hline \hline
	\multirow{2}{*}{TFI @ $h_c$}
	             & $2$ & $-3.233\ 573\ 421$ \\
	             & $3$ & $-3.234\ 260\ 711$ \\
	\hline \hline
	\multirow{4}{*}{$S = 1/2$ HB}
	             & $2$ & $-0.660\ 231\ 093$ \\
	             & $3$ & $-0.667\ 974\ 240$ \\
	             & $4$ & $-0.669\ 083\ 757$ \\
	             & $5$ & $-0.669\ 378\ 064$ \\
	\hline \hline
	\multirow{4}{*}{$S = 1/2$ XY}
	             & $2$ & $-0.547\ 658\ 559$ \\
	             & $3$ & $-0.548\ 706\ 183$ \\
	             & $4$ & $-0.548\ 810\ 284$ \\
	             & $5$ & $-0.548\ 818\ 968$ \\
	\hline
\end{tabular}

\caption{Best variational energies}
\label{tab:variational_energies}
\end{table}

\end{document}